\documentclass[10pt,twocolumn,twoside]{IEEEtran}    

\usepackage{amsmath}

\usepackage{amsfonts}

\usepackage{mathtools}

\usepackage{bbold}

\usepackage{upgreek}

\usepackage{graphicx}

\usepackage{subfigure}

\usepackage{bm}

\usepackage{cite}

\usepackage{color}

\usepackage{CJK}

\usepackage{verbatim}

\usepackage{url}

\usepackage{multirow}

\usepackage{array}

\usepackage{booktabs}

\usepackage{colortbl}

\usepackage{algorithm}

\usepackage{algorithmicx,algpseudocode}

\usepackage{diagbox}

\usepackage{makecell}

\usepackage{enumerate}

\usepackage{cases}

\usepackage{lipsum,cuted,arydshln}

\usepackage{threeparttable}

\usepackage{hyperref}
\hypersetup{colorlinks,citecolor=black,
            linkcolor=black,
            urlcolor=black}

\usepackage{varioref}

\usepackage{tablefootnote}

\usepackage[misc]{ifsym}

\usepackage{stfloats}

\usepackage{balance}

\usepackage{caption}

\usepackage[table]{xcolor}

\usepackage{setspace}

\usepackage{amssymb}
\usepackage{nomencl}
\makenomenclature

\usepackage{etoolbox}
\renewcommand\nomgroup[1]{%
  \item[\bfseries
  \ifstrequal{#1}{A}{Sets and Indices}{%
  \ifstrequal{#1}{B}{Parameters}{%
  \ifstrequal{#1}{C}{Variables}{}}}%
]}

\usepackage{pifont}

\def\QEDclosed{\mbox{\rule[0pt]{1.3ex}{1.3ex}}} 

\usepackage[thmmarks]{ntheorem}

\usepackage{nomencl}

\usepackage{wasysym}

\usepackage{hyperref}

\usepackage{algorithm}
\usepackage{algpseudocode}

\makeatletter
\def\UrlAlphabet{%
      \do\a\do\b\do\c\do\d\do\e\do\f\do\g\do\h\do\i\do\j%
      \do\k\do\l\do\m\do\n\do\o\do\p\do\q\do\r\do\s\do\t%
      \do\u\do\v\do\w\do\x\do\y\do\z\do\A\do\B\do\C\do\D%
      \do\E\do\F\do\G\do\H\do\I\do\J\do\K\do\L\do\M\do\N%
      \do\O\do\P\do\Q\do\R\do\S\do\T\do\U\do\V\do\W\do\X%
      \do\Y\do\Z}
\def\UrlDigits{\do\1\do\2\do\3\do\4\do\5\do\6\do\7\do\8\do\9\do\0}
\g@addto@macro{\UrlBreaks}{\UrlOrds}
\g@addto@macro{\UrlBreaks}{\UrlAlphabet}
\g@addto@macro{\UrlBreaks}{\UrlDigits}
\makeatother

{
  \theoremstyle{nonumberplain}
  \theoremheaderfont{\bfseries\em}
  \theorembodyfont{\normalfont}
  \theoremseparator{:}
  \theoremsymbol{\mbox{$\QEDclosed$}}
  \newtheorem{Proof}{Proof}
}

\theoremheaderfont{\bf\em}
\theorembodyfont{\normalfont}
\theoremseparator{:}

\newtheorem{Propos}{Proposition}

\newcommand* \cb[1]{\bm{#1}}

\definecolor{gray1}{rgb}{0.9,0.9,0.9}
\definecolor{gray2}{rgb}{0.8,0.8,0.8}
\definecolor{gray3}{rgb}{0.7,0.7,0.7}
\definecolor{white}{rgb}{1,1,1}

\begin{document}


\title{Detection-Triggered Recursive Impact Mitigation against Secondary False Data Injection Attacks in Microgrids}

\author{
\color{blue}Mengxiang Liu, Xin Zhang, Rui Zhang, Zhuoran Zhou, Zhenyong Zhang, and Ruilong Deng
\thanks{M. Liu, Z. Zhou, and X. Zhang are with the Department of Automatic Control and Systems Engineering, University of Sheffield, Sheffield, UK (e-mails: \{mengxiang.liu, zzhou106, xin.zhang1\}@sheffield.ac.uk).}
\thanks{R. Zhang is with the National Grid ESO, Wokingham, United Kingdom (e-mail: rui.zhang@nationalgrideso.com).}
\thanks{Z. Zhang and R. Deng are with the State Key Laboratory of Industrial Control Technology and the College of Control Science and Engineering, Zhejiang University, Hangzhou, China (e-mails: \{dengruilong, zhangzhenyong\}@zju.edu.cn).}

\thanks{{\color{black}This work was supported in part by the UK Research and Innovation Future Leaders Fellowship ‘Digitalisation of Electrical Power and Energy Systems Operation (DEEPS)’ under Grant MR/W011360/2 (Corresponding author: Xin Zhang)}.}
}
\maketitle
\begin{spacing}{1}
\begin{abstract}

The cybersecurity of microgrid has received widespread attentions due to the frequently reported attack accidents against distributed energy resource (DER) manufactures. Numerous impact mitigation schemes have been proposed to reduce or eliminate the impacts of false data injection attacks (FDIAs). Nevertheless, the existing methods either requires at least one neighboring trustworthy agent or may bring in unacceptable cost burdens. This paper aims to propose a detection-triggered recursive impact mitigation scheme that can timely and precisely counter the secondary FDIAs (SFDIAs) against the communication links among DERs. Once triggering attack alarms, the power line current readings will be utilised to observe the voltage bias injections through the physical interconnections among DERs, based on which the current bias injections can be recursively reconstructed from the residuals generated by unknown input observers (UIOs). The attack impacts are eliminated by subtracting the reconstructed bias from the incoming compromised data. The proposed mitigation method can work even in the worst case where all communication links are under SFDIAs and only require extra current sensors. The bias reconstruction performance under initial errors and system noises is theoretically analysed and the reconstruction error is proved to be bounded regardless of the electrical parameters. {\color{blue}To avoid deploying current sensors on all power lines, a cost-effective deployment strategy is presented to secure a spanning tree set of communication links that can guarantee the secondary control performance. Extensive validation studies are conducted in MATLAB/SIMULINK and hardware-in-the-loop (HIL) testbeds to validate the proposed method's effectiveness against single/multiple and continuous/discontinuous SFDIAs.}

\end{abstract}



\nomenclature[A, 01]{\(\mathcal{G}\)}{Set of generator buses}


\section{Introduction}

The recent cyberattack accidents against distributed energy resource (DER) manufactures like Vestas \cite{AccidentonVestas2021} and EnerCon \cite{AccidentonEnerCon2022} imply that the widely penetrated DERs have been frequently targeted by adversaries. The information forensics of these accidents further verified that the adversary's attempt is shifting from stealing sensitive data for ransom to interrupting DERs' normal operations. As one of the widely adopted forms that absorb DERs in the power grid, the microgrid has integrated massive advanced information and communications technologies like TCP/IP communication and CAN/Profinet to meet DERs' control and dispatch requirements. It is thus vital to foresee the potential cybersecurity issues in microgrids and take necessary cybersecurity enhancement strategies \cite{xu2020review}.

\begin{figure}[h]
  \centering
  \includegraphics[width=8cm]{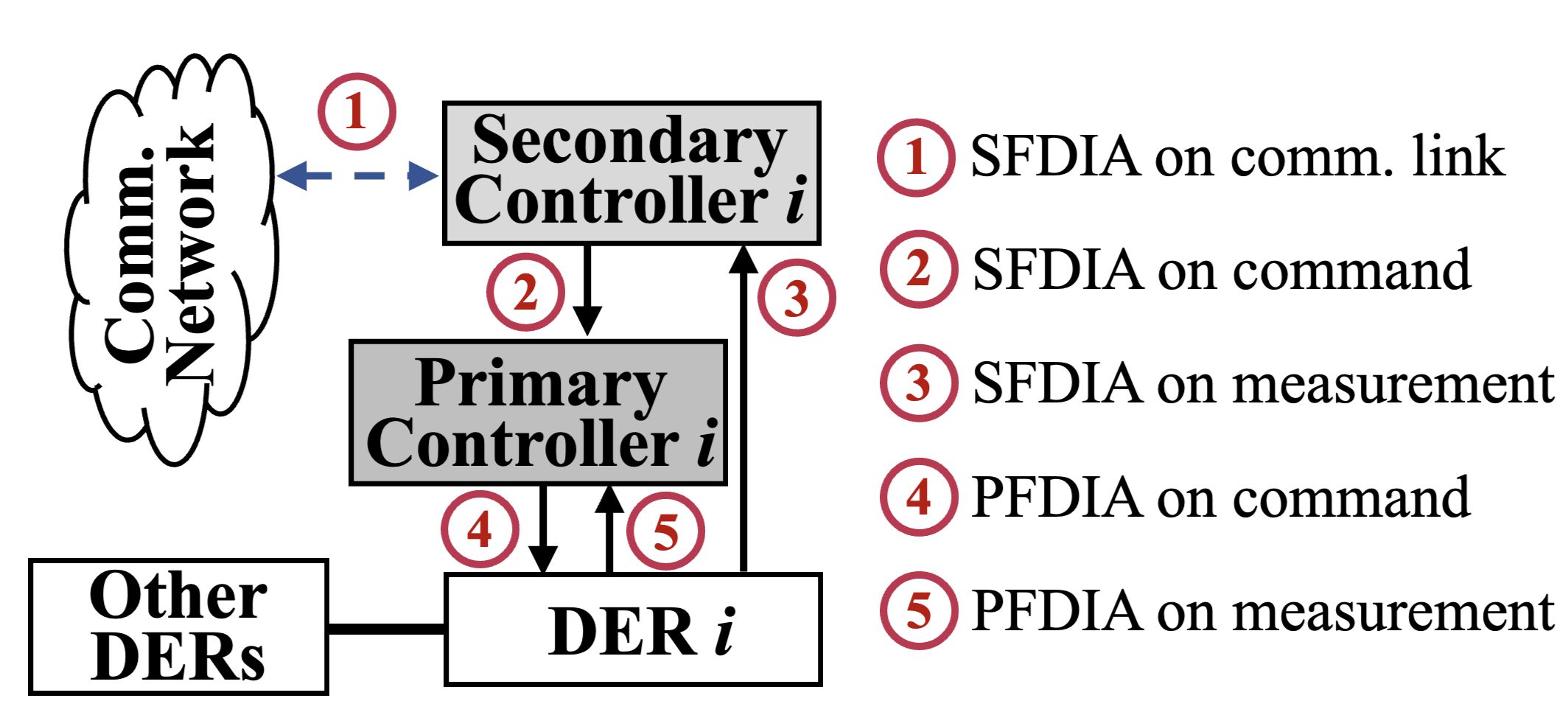}
  \captionsetup{font={footnotesize}}
  \caption{This figure illustrates the locations of five typical FDIAs against primary and secondary controllers in the microgrid.}\label{fig:AttackTypeIllustration}
\end{figure}

According to the hierarchical control framework of microgrid as shown in Fig. \ref{fig:AttackTypeIllustration}, where the primary controller regulates the local DER states and the secondary controller coordinates DERs to achieve global objectives like load sharing \cite{yang2018hierarchical}, five typical false data injection attacks (FDIAs) against the two controllers by exploiting protocol vulnerabilities \cite{8918111}, firmware rootkit \cite{garcia2017hey}, and sensor spoofing tools \cite{barua2020hall} can be identified. These secondary and primary FDIAs (SFDIA and PFDIA) can disrupt the normal behaviours of DERs, initiate protective actions, and potentially lead to cascading failures. As a result, it is necessary to make appropriate cyber-resilient prevention, detection, mitigation, and recovery strategies \cite{liu2023enhancing}. This paper mainly focuses on the during-attack impact mitigation phase that is used to reduce or eliminate the attack impacts such that there would be enough time to make the cyber recovery plan.

Numerous literature has emerged in the topic of designing cyber-resilient impact mitigation strategies for the microgrid under FDIAs, which can be generally classified as model-based and data-driven according to the principles of adopted methods. Focusing on the SFDIA on control commands (\ding{173}), Zuo \textit{et al.} \cite{zuo2020distributed} proposed a resilient cooperative secondary controller that can mitigate the impacts of bounded and unbounded bias injections in an adaptive manner. Given the SFDIAs (\ding{173}) in energy storage dominated micorgrids \cite{9039721} and networked AC/DC microgrids \cite{9184978}, Wang and Deng \textit{et al.} designed distributed resilient secondary controllers by introducing adaptive techniques to maintain the overall control objectives. 
To infer the bias injected into the communication links, Jin \textit{et al.} \cite{jin2022distributed} aimed to analyse the impacts of SFDIAs (\ding{172}) on voltages to obtain the exact bias expression. But this method has limitations in handling the case where multiple SFDIAs with general bias types exist. In response to the SFDIAs on communication links (\ding{172}) and control commands (\ding{173}), Sahoo \textit{et al.} proposed to reconstruct an detection-triggered secondary control error signal using the trustworthy agents' information \cite{sahoo2020multilayer,rath2023self}. This mitigation strategy requires that at least one of the neighboring links/agents should be trustworthy to guarantee the credibility of reconstructed signals.

Considering the PFDIA on the control command (\ding{175}), Sadabadi \textit{et al.} \cite{9600614} introduced a resilient voltage tracking and current sharing primary controller to mitigate the potential adverse impacts. When the secondary communication links (\ding{172}), control commands (\ding{173}), and measurements (\ding{174}) are all subject to unbounded FDIAs, Zuo \textit{et al.} \cite{zuo2020resilient} introduced a hidden secure communication network where the transmitted data is immune to malicious modifications to correct the hazardous control actions. With the provision of an internet-of-things (IoT) based digital twin that interacts with the microgrid control system, Saad \textit{et al.} \cite{saad2020implementation} developed practical resilient control algorithms by replacing the malicious data with those received from the Luenberger observer based IoT shadow. The hidden network and digital twin based mitigation strategies can effectively maintain the microgrid control performance even under hybrid SFDIAs (\ding{172}, \ding{174}) and PFDIAs (\ding{176}), but the associated extra costs may be unacceptable for the system operator.

The potential of data-driven methods in mitigating FDIAs' impacts within microgrids has also been shed initial lights recently. By training an artificial neural network (ANN) with the inputs of DC-DC converters' output voltages and currents, Habibi \textit{et al.} \cite{habibi2020false} proposed a novel data-driven method to mitigate the impacts of the SFDIA compromising the parallel converters' point of common coupling (PCC) voltage readings (\ding{174}). To enable its decentralized implementation, Habibi \textit{et al.} \cite{habibi2021decentralized} successfully designed an ANN to recover the converter's legitimate output current under FDIAs using the inputs of renewable source's output current and primary voltage control signal. To alleviate the necessity of multiple training processes for a well-tuned deep neural network based estimator, which is used to recover the sum of communicated signals using the local measurements, Xia \textit{et al.} \cite{10210203} designed a representation subspace distance based transfer learning model. 

\begin{table}[!h]
    \small
    \caption{Comparisons between the proposed impact mitigation method and the existing ones against SFDIAs}
    \label{tab:comparative table with literature}
    \centering
    \begin{threeparttable}
    \begin{tabular}{p{1cm}<{\centering}p{1.5cm}<{\centering}p{0.6cm}<{\centering}p{0.4cm}<{\centering}p{0.4cm}<{\centering}p{2.1cm}<{\centering}}
        \toprule
        \multirow{2}{1cm}{\centering \textbf{Category}} & \multirow{2}{1.5cm}{\centering \textbf{Literature}} &  \multicolumn{3}{c}{ \textbf{SFDIA Types}} & \\
         & & \textcolor{black}{\ding{172}} &  \textcolor{black}{\ding{173}} & \textcolor{black}{\ding{174}} & \multirow{-2}*{ \centering \textbf{Extra Cost}} \\ \midrule
        \multirow{5}{0.8cm}{Model-based} & \cite{zuo2020distributed,9039721,9184978} & \ding{55} & \ding{51}  & \ding{55} & No \\
        & \cite{jin2022distributed} & \ \cellcolor{gray1} \ding{51}\textcolor{black}{$^{\circ}$} & \ding{55} & \ding{55}  &  No \\
        & \cite{sahoo2020multilayer,rath2023self} & \ \cellcolor{gray1} \ding{51}\textcolor{black}{$^*$} & \ding{51} & \ding{55}  & Blockchain \\
         & \cite{zuo2020resilient} & \ding{51} & \ding{51} & \ding{51} & \cellcolor{gray1} Hidden layer \\ 
         & \cite{saad2020implementation} & \ding{51} & \ding{55} & \ding{51} & \cellcolor{gray1} Digital twin \\ \midrule
         \multirow{2}{0.8cm}{Data-driven} & \cite{habibi2020false,habibi2021decentralized} & \ding{55} & \ding{55} &  \ding{51} & No  \\
         & \cite{10210203} & \ding{51} & \ding{55} & \ding{55} & No \\ \midrule
          \multicolumn{2}{c}{\cellcolor{gray3}\centering \textcolor{blue}{\textbf{Model-based Our work}}} & \cellcolor{gray3}\textcolor{blue}{\ding{51}} & \cellcolor{gray3} \ding{55} & \cellcolor{gray3} \ding{55} & \cellcolor{gray3} \textcolor{blue}{\textbf{Current sensor}} \\ \bottomrule
    \end{tabular}
    \begin{tablenotes}
  {\footnotesize
  \item[\textcolor{black}{$\circ$}] \textcolor{black}{The number of SFDIA is limited by $1$.}
  \item[\textcolor{black}{*}] \textcolor{black}{There should be at least one trustworthy beighboring agent.}}
\end{tablenotes}
\end{threeparttable}
\end{table}

A tabulated illustration of the reviewed impact mitigation methods against SFDIAs is provided in TABLE \ref{tab:comparative table with literature}. First of all, the investigation of the impact mitigation schemes \cite{zuo2020distributed,9039721,9184978} against the SFDIA on control commands (\ding{173}) has been well developed and are quite mature. Nevertheless, existing impact mitigation schemes against the SFDIA on communication links (\ding{172}) either have assumptions on the number of compromised links \cite{jin2022distributed,sahoo2020multilayer,rath2023self} or may bring in unacceptable cost burdens for the system operator \cite{zuo2020resilient,saad2020implementation}. While the data-driven mitigation methods \cite{habibi2020false,habibi2021decentralized,10210203} have not incorporated the physical knowledge, under which the reconstructed data may not comply with the underlying physical dynamics and thus the reconstructed data can oppositely exacerbate the attack impacts. 

\begin{figure*}[t]
  \centering
  \includegraphics[width=18cm]{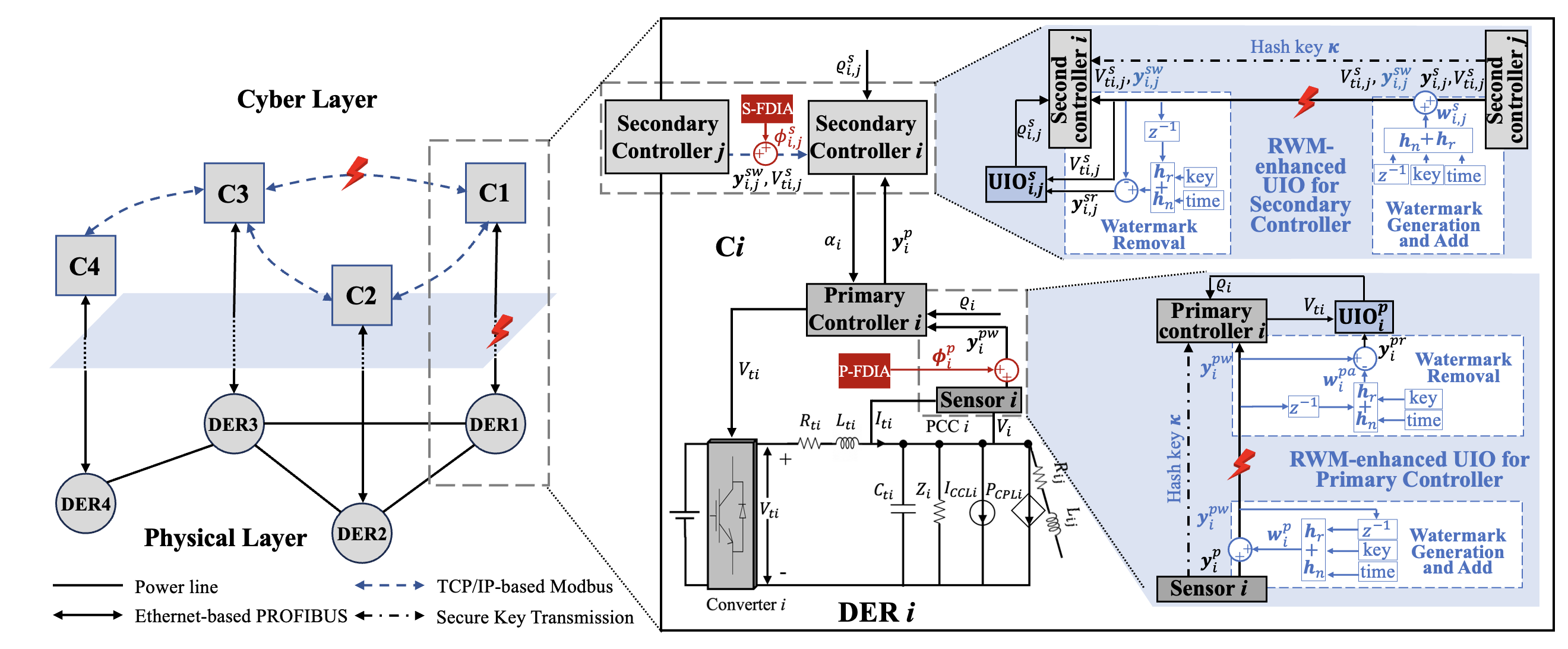}
  \captionsetup{font={footnotesize}}
  \caption{\color{blue}The left part of the figure depicts the highly abstracted cyber and physical framework of microgrids, the middle part illustrates the specific cyber-physical couplings and exposed attack surfaces to SFDIAs, and the right part shows the graphical framework of the detection-triggered recursive bias reconstruction and impact mitigation scheme and its integration with original secondary controllers.}\label{fig:system_model}
\end{figure*}

Towards this end, this paper aims to design an effective and cost-efficient impact mitigation scheme against the SFDIA on communication links (\ding{172}) based on the unknown input observer (UIO), which has been proved useful in detecting random \cite{gallo2018distributed} and stealthy \cite{gallo2020distributed,9621221,9815319} FDIAs. As shown in Fig. \ref{fig:system_model}, the impact mitigation scheme will only be activated after UIOs have perceived anomalous behaviours. We propose a novel idea to firstly observe the voltage bias injection through incorporating the physical interconnection among DERs, i.e., utilising the power line current to observe the voltage of interconnected DER.
Based on the observed voltage bias injection, the current bias injections could then be recursively reconstructed from the UIO's detection residuals. Finally, the reconstructed biases are subtracted from the compromised communicated data to mitigate the attack impacts. The main contributions of this paper are:

\begin{itemize}
    \item We propose an innovative detection-triggered recursive impact mitigation scheme against the SFDIA on communication links (\ding{172}), which is effective even when all communication links are compromised and requires only to install additional current sensors on power lines.
    \item We derive the explicit relations between UIO residuals and bias injections, based on which a recursive bias reconstruction scheme is intuitively established. Moreover, we prove that the reconstruction error is bounded by a term determined the system noises' bounds regardless of the electrical parameters.
    \item {\color{blue}We present a cost-effective deployment strategy for current sensors based on existing measuring infrastructure, aiming to secure a spanning tree set of communication links that can guarantee the secondary control performance.
    \item We conduct extensive validation studies in MATLAB/SIMULINK and hardware-in-the-loop (HIL) testbeds to illustrate the proposed method's effectiveness against various SFDIAs in the presence of load variations, DERs plugging-in, and voltage fluctuations.}
\end{itemize}

\section{Isolated Microgrid System Model}
{\color{blue}As shown in Fig. \ref{fig:system_model}, we consider an isolated DC microgrid consisting of $N\ge 2$ DERs, where renewable energy resources (RESs) including photovoltaic (PV) cells are operated with sufficient storage units to eliminate the intermittent power generation resulting from weather variations and therefore can be simplified as a constant voltage source \cite{yu2023coordinated,han2018stability}.} The DC-DC converter is commanded to supply the local ZIP load through an RLC filter. The electrical network that physically connects the DERs is denoted by a weighted bidirectional graph $\mathcal{G}_{el} = \{ \mathcal{A}, \mathcal{E}_{el} \}$, where the set $\mathcal{A}$ collects all the DER nodes and the set $\mathcal{E}_{el}$ includes the power lines that connect the DERs with the set of weights as the corresponding line conductance, that is, $\frac{1}{R_{ij}}, \forall (i,j) \in \mathcal{E}_{el}$.\footnote{The inductive part, i.e., $L_{ij}$ as shown in Fig. \ref{fig:system_model}, is omitted as it is usually far smaller than the resistive part.} The set that contains the physical neighbours of the DER $i \in \mathcal{A}$ in graph $\mathcal{G}_{el}$ is represented by $\mathcal{N}_i^{el}$. The communication network that transmits data between DERs is also indicated by a weighted bidirectional graph $\mathcal{G}_c = \{ \mathcal{A}, \mathcal{E}_c\}$, where the set $\mathcal{E}_c$ gathers all communication links with the weights designed as $a_{ij}^c, \forall (i,j) \in \mathcal{E}_c $. The cyber neighbors of DER $i\in\mathcal{A}$ in graph $\mathcal{G}_c$ are collected in the set $\mathcal{N}_i^c$. We assume that the electrical and communication graphs share the same topology are both connected, which establish the sufficient conditions for achieving stable control \cite{tucci2018stable}.

Since the microgrid is operated close to the nominal reference point of common coupling (PCC) voltage $V_{ref,i}$, the constant power load (CPL) $P_{CPL,i}$ is linearised around this nominal operating point as
\begin{align}\label{eq: CPL linearisition}
    I_{CPL,i} = -\frac{P_{CPL,i}}{V_{ref,i}^2}V_i + 2\frac{P_{CPL,i}}{V_{ref,i}},
\end{align}where the former part is a negative impedance related to the actual PCC voltage $V_i$ and the latter part is a constant current. Under linearisation \eqref{eq: CPL linearisition}, the original ZIP load can be characterised by the equivalent constant impedance ($Z_{Li}$) and current ($I_{Li}$) loads as follows:
\begin{align}
    \frac{1}{Z_{Li}} = \frac{1}{Z_i} - \frac{P_{CPL,i}}{V_{ref,i}^2} & \label{eq: linearised ZIP load 1},\\
    I_{Li} = I_{CCL,i} + 2\frac{P_{CPL,i}}{V_{ref,i}} & \label{eq: linearised ZIP load 2},
\end{align}where $Z_i$ and $I_{CCL,i}$ denote the original constant impedance and current parts of the ZIP load, respectively. Based on the linearised ZIP load \eqref{eq: linearised ZIP load 1}-\eqref{eq: linearised ZIP load 2}, the dynamics of the RLC filter inside DER $i$ can be obtained after utilising the Kirchhoff voltage and current laws and incorporating the quasi-stationary line approximation ($L_{ij} \approx 0$), i.e., 
{
\footnotesize
\begin{align}\label{eq:DER model}
\left\{
\begin{aligned}
\frac{d V_i}{dt}=\frac{1}{C_{ti}}I_{ti}+&\sum_{j\in\mathcal{N}_i^{el}}\frac{1}{C_{ti}R_{ij}}(V_j-V_i)-\frac{1}{C_{ti}}(I_{Li}+\frac{V_i}{Z_{Li}})\\
\frac{dI_{ti}}{dt}=-\frac{1}{L_{ti}}V_i&-\frac{R_{ti}}{L_{ti}}I_{ti}+\frac{1}{L_{ti}}V_{ti}
\end{aligned}
\right.,
\end{align}}where $V_{ti}$ denotes the converter's output voltage, $I_{ti}$ represents the output current from the renewable source, and $R_{ti}, L_{ti}$, and $C_{ti}$ are the RLC filter parameters. Denote the system state by $\cb{x}_i = [V_i, I_{ti}]^{\rm T}$, the system dynamics \eqref{eq:DER model} can be rewritten as
\begin{align}\label{eq: continuous system dynamics}
    \dot{\cb{x}}_i(t) = A_{ii} \cb{x}_i(t) + \cb{b}_i u_i(t) + \cb{m}_i d_i(t),
\end{align}where $A_{ii}, \cb{b}_i$, and $\cb{m}_i$ denote the system matrix, primary input vector, and unknown input vector, respectively, and are detailed in Appendix \ref{appendix: system parameters}. The primary input $u_i(t) = V_{ti} (t)$ represents the voltage regulation command to the DC-DC converter and the unknown input $d_i(t) = I_{Li} + \sum_{j\in\mathcal{N}_i^{el}} -\frac{1}{R_{ij}} V_j(t)$ includes the current load as well as the physical couplings with neighboring DERs. 

To be compatible with the digital signal based processor, the dynamical model \eqref{eq: continuous system dynamics} is discretized with sampling time $T_{samp}$ and is blended with fully measured system states along with bounded process and measurement noises as the following state-space model:
{
\begin{align}\label{eq: discrete-time DER model with measurement and noise}
\left\{
\begin{aligned}
{\cb{x}}_i(k+1) &= A_{ii}^d \cb{x}_i(k) + \cb{b}_i^d u_i(k) + \cb{m}_i^d d_i(k) + \cb{\omega}_i(k) \\
\cb{y}_i(k) &= \cb{x}_i(k) + \cb{\rho}_i(k)
\end{aligned}
\right.,
\end{align}}where system parameters $A_{ii}^d, \cb{b}_i^d$, and $\cb{m}_i^d$ are derived from the original parameters in \eqref{eq: continuous system dynamics} with sampling time $T_{samp}$ and their relations are detailed in appendix \ref{appendix: system parameters}. The system states are fully measured as local output vector $\cb{y}_i(k)$, and bounded process and measurement noises satisfy $|\cb{\omega}_i(k)| \le \bar{\cb{\omega}}_i$ and $|\cb{\rho}_i(k)| \le \bar{\cb{\rho}}_i$, respectively.

To achieve the reference PCC voltage tracking, the primary control input $u_i(k)$ is calculated based on the general Proportion-Integral algorithm as follows
{
\small
\begin{align}\label{eq: primary control input}
    u_i(k) = \big(\cb{g}_i^P\big)^{\rm T} \cb{y}_i^p(k) + g_i^I\sum_{l=0}^k \Big(V_{ref,i} + \alpha_i (l) - \cb{\kappa}^{\rm T} \cb{y}_i^p (l)\Big),
\end{align}}where $\cb{g}_i^P$ and ${g}_i^I$ contain proportional and integral control gains, respectively, $\cb{y}_i^p(k)$ is the local outputs available to the primary controller that will be equal to $\cb{y}_i(k)$ in the normal case, and $\alpha_i (k)$ denotes the secondary control input that is computed utilising the data transmitted over the communication network. {\color{black}To satisfy the hard real-time requirement of primary control, field-level communication protocols like CAN and Profinet are usually adopted to establish the sensor channel.} Constant vector $\cb{\kappa} = [1,0]^{\rm T}$ is used to extract the PCC voltage information for the calculation of accumulated voltage tracking errors. 

The secondary controller is deployed on top of the primary controller to regulate the voltage tracking objective in a coordinated manner to achieve the overall control objectives in the microgrid including voltage balancing and current sharing. {\color{blue}TCP Modbus communication protocol is recommended to accomplish the data exchange between DERs.} In the essential principle of consensus control, the secondary control input is calculated as
\begin{align}\label{eq: Secondary control input}
    \alpha_i(k) = \cb{\iota}^{\rm T}\sum_{l=0}^k\sum_{j\in\mathcal{N}_i^c}a_{ij}^c (\frac{\cb{y}_{i,j}^s(l)}{I_{tj}^s} - \frac{\cb{y}_{i}^p(l)}{I_{ti}^s}),
\end{align}where $\cb{y}_{i,j}^s(k)$ denotes the output vector of DER $j\in \mathcal{N}_i^c$ transmitted over link $(i,j)\in\mathcal{E}_c$ that will be equal to $\cb{y}_j^p(k)$ in the normal case, $I_{tj}^s >0$ and $I_{ti}^s >0$ are the rated output currents corresponding to DERs $j$ and $i$, respectively, and constant vector $\cb{\iota} = [0,1]^{\rm T}$ is applied to obtain the accumulated current sharing errors.

\begin{figure}[!h]
    \centering
    \includegraphics[width = 9cm]{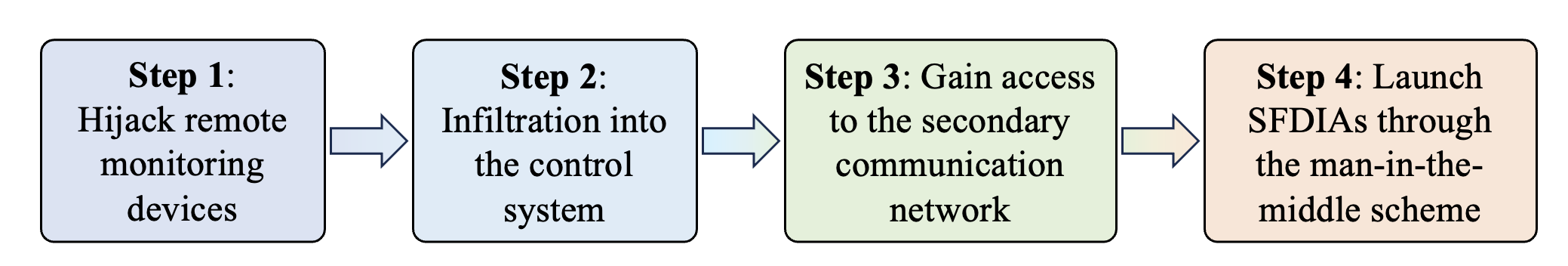}
    \captionsetup{font={footnotesize}}
    \caption{\color{blue}This figure illustrates a practical attack path consisting of four key steps for launching the SFDIA in microgrids.}
    \label{fig:man-in-the-middle}
\end{figure}

\subsection{Attack Model}
{\color{blue}For the data interaction required by secondary control, the TCP Modbus communication protocol has been widely recommended owing to its open standard and easy-to-use characteristic that facilitate the interoperability among devices from different manufacturers \cite{liu2023enhancing}. The insufficient security mechanisms of Modbus in data encryption, node authentication, and key management have exposed the secondary communication network to various cyber attacks like FDIAs \cite{9815319}, which can be launched through the well-known man-in-the-middle attack scheme once the adversary gains access to the network \cite{pu2024cormand2}. According to the recently disclosed attack incident against solar power facilities in Japan, the malicious adversary hijacked 800 SolarView Compact remote monitoring devices by exploiting a known flaw that was not patched to engage in bank account thefts \cite{SolarAttackJapan2024}. Utilising the hijacked remote monitoring device as a springboard, the adversary could further infiltrate into the control system and severely affect the operation of solar facilities. Inspired by this, a practical attack path demonstrating the steps of launching the SFDIA in microgrids is shown in Fig. \ref{fig:man-in-the-middle}. The first step is to hijack remote monitoring devices by exploiting zero-day or disclosed flaws. Conducting infiltration from hijacked remote monitoring devices into the control system, the adversary makes lateral movements to the monitoring host of DER in step 2, after which the adversary would be able to access the secondary control's communication network in step 3. Finally, in step 4, the adversary can launch SFDIAs through the man-in-the-middle scheme, where the malicious entity is inserted between the two legal communication DERs to modify the transmitted data packets while deceiving the two DERs via ARP spoofing \cite{pu2024cormand2}.}

In DER $i$, the SFDIA that target at its secondary communication links is modelled as
\begin{align}\label{eq: FDIA against secondary controller}
    {\rm{SFDIA:}}\ \ \ \   
    \cb{y}_{i,j}^s (k) &= \cb{y}_j^p (k) + \cb{\phi}_{i,j}^{sy} (k), 
\end{align}where $\cb{\phi}_{i,j}^{sy}$ denotes the bias vector injected into neighboring DER $j$'s transmitted output vector. 
The design of $\cb{\phi}_{i,j}^{sy}, \forall i,j\in\mathcal{A}$ against multiple communication links can be implemented in the either independent or conspiring manner to incur hazardous consequences including voltage instability \cite{9484453} and economic loss \cite{9815319}. Moreover, we consider the worst case where the adversary's resources are not restricted and arbitrary number of communication links can be compromised.

\subsection{Unknown Input Observer based Attack 
Detection}
The UIO is originally designed to estimate system states in the presence of the unknown inputs resulted from uncertain system parameters and unexpected component failures, under which performance guaranteed controllers can be designed to tolerate these uncertainties \cite{chen1996design}. Recently, the potential application of UIO to detecting FDIAs in large-scale interconnected systems like microgrids has been well investigated \cite{gallo2018distributed}. The basic idea is to use the estimated system states to measure if the trajectory of received data satisfy DER system dynamics \eqref{eq: discrete-time DER model with measurement and noise}. Specifically, the UIO$_{i,j}^s$ designed for the DER $j$'s data communicated through the communication link is represented as
{
\small
\begin{align}\label{eq: UIO for secondary controller}
&{\rm UIO}_{i,j}^s: \nonumber \\
&\left\{
\begin{aligned}
\cb{z}_{i,j}^{s} (k+1) &= F_j^s \cb{z}_{i,j}^s (k) + T_j^s \cb{b}_j^d u_{i,j}^s(k) + \hat{K}_j^s \cb{y}_{i,j}^s (k) \\
\hat{\cb{x}}_{i,j}^{s} (k) &= \cb{z}_{i,j}^{s} (k) + H_j^s \cb{y}_{i,j}^{s} (k)
\end{aligned}
\right.,
\end{align}}where $\cb{z}_{i,j}^s$ is UIO$_{i,j}^s$'s internal state vector, $\hat{\cb{x}}_{i,j}^s$ is the estimated DER $j$'s state vector, and $u_{i,j}^s = u_j$ is the communicated DER $j$'s converter command. The UIO matrix $F_j^s$ is designed to be stable, i.e., the real parts of its eigenvalues need to satisfy $|\lambda_l^{\rm re}(F_j^s)| < 1, \forall l \in \{1,2\}$, to converge the state estimation error and matrix $T_j^s$ is designed according to $T_j^s\cb{m}_j^d = \mathbb{0}^{2\times1}$ to eliminate the impacts of unknown inputs. {\color{blue}The principles of choosing UIO matrices $F_j^s, T_j^s, \hat{K}_j^s, H_j^s$ should satisfy
\begin{align}
    &T_j^s = {\rm I}^2 - H_j^s \\
    &T_j^s\cb{m}_j^d = \mathbb{0}^{2\times1} \label{eq: UIOSec_2} \\
    &\hat{K}_j^s = K_{j1}^s + K_{j2}^s \label{eq: UIOSec_3} \\
    &F_j^s = T_j^s A_{jj}^d - K_{j1}^s \label{eq: UIOSec_4} \\
    &K_{j2}^s = F_j^sH_j^s \label{eq: UIOSec_5}
\end{align} to eliminate the impact of unknown inputs and initial estimation errors on the state estimation accuracy.} The detection residual is derived by comparing the estimated state vector with the actual output vector as 
\begin{align}\label{eq: Residual calculation}
    \cb{r}_{i,j}^s (k) = \cb{y}_{i,j}^s (k) - \hat{\cb{x}}_{i,j}^s(k)
\end{align}
and in the normal operation its analytical expression can be obtained considering \eqref{eq: discrete-time DER model with measurement and noise} and \eqref{eq: UIO for secondary controller}:
\begin{align}\label{eq: primary residual}
    &\cb{r}_{i,j}^s(k)=(F_j^s)^k\big(\cb{r}_{i,j}^s(0)-T_j^s\cb{\rho}_j(0)\big)+T_{j}^s\cb{\rho}_j(k) \nonumber\\
    &\ \ \ \ \ \ \ \ \ \sum_{l=0}^{k-1}\big(F_j^s\big)^{k-1-l}\big(T_j^s\cb{\omega}_j(l)-\hat{K}_j^s\cb{\rho}_j(l)\big),
\end{align}where $\cb{r}_{i,j}^s(0)$ is the initial residual vector and will be equal to zero after strategically choosing the initial UIO state vector as $\cb{z}_{i,j}^s(0) = T_j^s\cb{y}_{i,j}^s(0)$. Given the bounded system noises, the upper bounded of $|\cb{r}_{i,j}^s|$ can be thus calculated as 
\begin{align}\label{eq: primary residual upper bound}
    |\cb{r}_{i,j}^s(k)| \le &\bar{\cb{r}}_{i,j}^s (k) = \nu_j^s(\varsigma_j^s)^k\big|T_j^s\big|\bar{\cb{\rho}}_j+|T_j^s|\bar{\cb{\rho}}_j \nonumber\\
    &\sum_{l=0}^{k-1}\nu_j^s(\varsigma_j^s)^{k-1-l}\big(|T_j^s|\bar{\cb{\omega}}_j+|\hat{K}_j^s|\bar{\cb{\rho}}_j\big),
\end{align}where positive scalars $\nu_j^s$ and $0 < \varsigma_j^s <1$ are chosen such that $||(F_j^s)^k|| \le \nu_j^s(\varsigma_j^s)^k$.


Once either of the following conditions is satisfied, attack alarms will be triggered for the corresponding data
\begin{align}\label{eq: detection conditions under SFDIA}
    r_{i,j}^{s,V} (k) > \bar{r}_{i,j}^{s,V} (k)\ \ &{\rm or}\ \ r_{i,j}^{s,I} (k) > \bar{r}_{i,j}^{s,I} (k), 
\end{align}where $\cb{r}_{i,j}^s = [r_{i,j}^{s,V}, r_{i,j}^{s,I}]^{\rm T}$ and $\bar{\cb{r}}_{i,j}^s = [\bar{r}_{i,j}^{s,V}, \bar{r}_{i,j}^{s,I}]^{\rm T}$. The attack alarm $\varrho_{i,j}^s = 1$ indicate that the neighboring communicated measurement output $\cb{y}_{i,j}^s$ are encountering SFDIAs. 

\begin{figure*}[tb]
    \centering
    \includegraphics[width = 18cm]{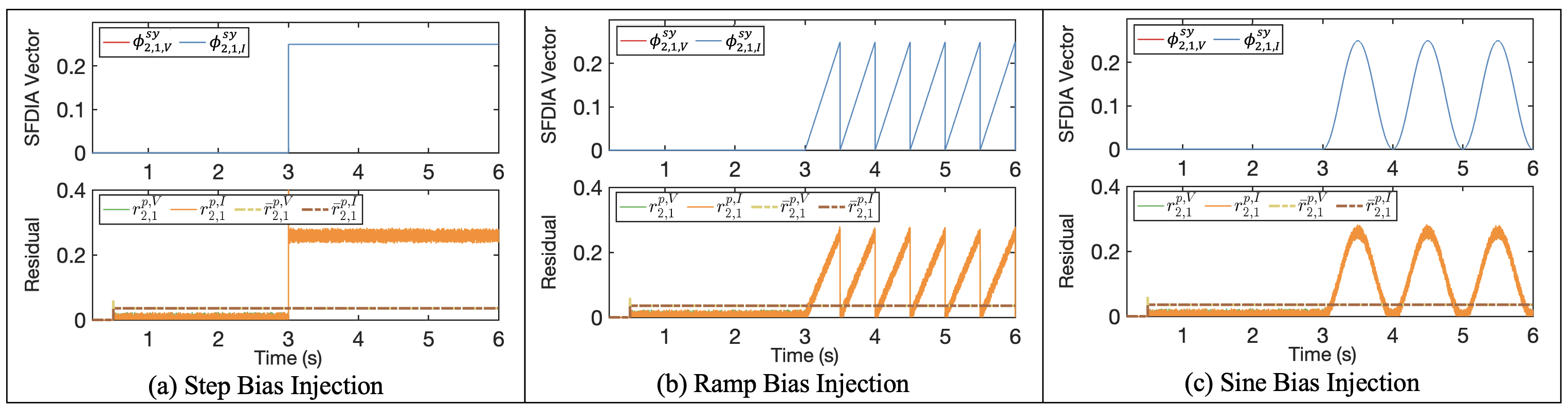}
    \captionsetup{font={footnotesize}}
    \caption{\color{blue}Motivating examples indicating the relations of UIOs' residuals and Step, Ramp, and Sine bias injections.}
    \label{fig:RelationsofResidualsandBiasInjections}
\end{figure*}

\section{Motivating Example and Problem Formulation}
It is not the end for the cyber-resilient operation of microgrids to perceive the existence, type, and location of unpredictable attack events. It is equivalently important to adopt appropriate countermeasure actions to mitigate and even eliminate the adversarial impacts of attacks after successfully detecting them. The detectability of UIOs against both the random \cite{gallo2018distributed} and stealthy \cite{gallo2020distributed,9621221,9815319} FDIAs in microgrids has been thoroughly investigated. It has been well known to utilise the residuals generated by UIOs \eqref{eq: primary residual} to validate the adherence of the received signal's trajectory to the underlying DER dynamics. \textit{However, a hidden and usually ignored fact underlying the generated residuals is that they also synthesize the bias injection information}. Hence, in addition to attack detection, another functionality of UIOs' residuals may be to reconstruct the injected biases, which can be then used to recover the legitimate data for the mitigation of attack impacts. 

To intuitively reflect the relations between UIOs' residuals and the SFDIA biases, three motivating examples are provided. The four-node DC microgrid with the cyber and physical typologies shown in Fig. \ref{fig:system_model} is established, where the communication link between DERs $1$ and $2$ is subject to SFDIAs and three representative types of bias injections (Step, Ramp, and Sine) into the neighboring output vector, i.e., $\cb{\phi}_{2,1}^{sy} = [\phi_{2,1,V}^{sy},\phi_{2,1,I}^{sy}]^{\rm T}$, are considered. {\color{blue}When picturing the bias injections and corresponding residuals in Fig. \ref{fig:RelationsofResidualsandBiasInjections}, an intuitive observation is that the residual trajectory can reflect many features of the injected biases like their steady-state values and changing patterns. Specifically, under the step bias injection case in sub-figure (a), the steady-state bias injections will result in steady-state increment on the corresponding residuals. Moreover, when the bias injections possess changing patterns like ramp and sine signals in sub-figures (b) and (c), the same patterns can be observed from the resulted residuals.} In addition to that, there are explicit theoretical links between residuals and the associated biases as reflected in \eqref{eq: discrete-time DER model with measurement and noise} and \eqref{eq: UIO for secondary controller}, which provide theory basis for the forthcoming reconstruction of SFDIA biases.

Therefore, based on the attack alarms triggered by UIOs, this paper aims to propose a post-detection bias injection reconstruction and impact mitigation scheme based on the residuals generated by UIOs. Three essential problems need to be sorted out: 1) Investigation of the analytical relations between residuals and the corresponding bias injections, based on which an computation-friendly and cost-efficient bias injection reconstruction scheme will be designed. 2) Analysis of the impact of initially estimated inaccurate bias injections on the subsequently reconstructed bias injections since it is hard to know the exact bias injections at the alarm flagging time. 3) Performance analysis of proposed bias reconstruction scheme in the presence of bounded system noises as small fluctuations can still significantly diverge the reconstructed biases from real ones. {\color{blue}4) Cost-effective deployment of current sensors that can ensure the secondary control performance under SFDIAs with minimal number of required sensors.}

\begin{figure}[h]
  \centering
  \includegraphics[width=9cm]{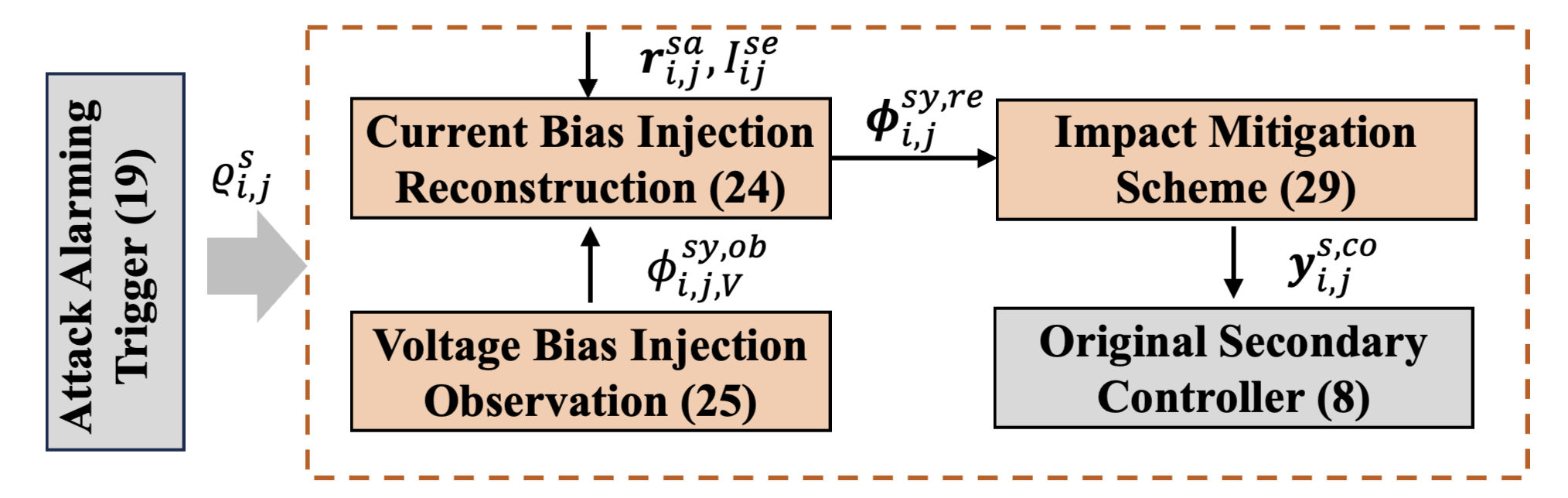}
  \captionsetup{font={footnotesize}}
  \caption{This figure elaborates the working flow of the detection-triggered bias reconstruction and impact mitigation scheme.}\label{fig:biasinjectionandimpactmitigationscheme}
\end{figure}

\section{Detection-Triggered Bias Reconstruction and Impact Mitigation}
This section introduces the theoretical links between UIO residuals and SFDIA bias injections, based on which the bias reconstruction and impact mitigation scheme are detailed along with rigorous performance analysis. The working flow of the proposed scheme is shown in Fig. \ref{fig:biasinjectionandimpactmitigationscheme}, where the attack alarming trigger generated by \eqref{eq: detection conditions under SFDIA} will activate the voltage bias observation action \eqref{eq: Observed SFDIA voltage bias} and current bias reconstruction function \eqref{eq: SFDIA bias reconstruction}. Then, the reconstructed bias SFDIA vector $\cb{\phi}_{i,j}^{sy,re}$ will be fed back to the impact mitigation scheme \eqref{eq: SFDIA imapct mitigation} to eliminate the attack impact on the secondary controller \eqref{eq: Secondary control input}.

\subsection{Explicit Relations between Residuals and SFDIA Biases}

The explicit relations between residual $\cb{r}_{i,j}^{sa}$ and bias injection $\cb{\phi}_{i,j}^{sy}$ can be derived from the DER dynamics \eqref{eq: discrete-time DER model with measurement and noise}, SFDIA model \eqref{eq: FDIA against secondary controller}, and UIO dynamics \eqref{eq: UIO for secondary controller}. According to the bias injection forms \eqref{eq: FDIA against secondary controller} and UIO$_{i,j}^s$ \eqref{eq: UIO for secondary controller}, the residual under attacks $\cb{r}_{i,j}^{sa}(k+1)$ can be calculated:
\begin{align}
\cb{r}_{i,j}^{sa}(k+1)  &= \cb{y}_{i,j}^s (k+1) - \hat{\cb{x}}_{i,j}^{s}(k+1)  \label{eq: Res Calculation1} \\
& = T_{j}^s\cb{y}_{i,j}^s(k+1) - \cb{z}_{i,j}^{s} (k+1) \nonumber \\
& = T_j^sA_{jj}^d\cb{y}_j(k+1) - F_j^s\cb{z}_{i,j}^s(k) + \nonumber \\
&-T_j^s\cb{b}_j^du_{i,j}^s(k) - \hat{K}_j^s\cb{y}_{i,j}^s(k) + T_j^s\cb{\phi}_{i,j}^{sy}(k+1). \nonumber
\end{align}
The DER $j$'s dynamics similar as \eqref{eq: discrete-time DER model with measurement and noise} can be used to expand $\cb{y}_j(k+1)$ and along with \eqref{eq: UIOSec_2}, the above expression \eqref{eq: Res Calculation1} is transformed to
\begin{align}
    \cb{r}_{i,j}^{sa}(k+1)  &= T_j^sA_{jj}^d\cb{x}_j(k) + (F_j^sH_j^s-\hat{K}_j^s)\cb{y}_{i,j}^s(k) +  \label{eq: Res Calculation2} \\ 
    &-F_j^s\hat{\cb{x}}_{i,j}^s(k) + T_j^s\cb{\phi}_{i,j}^{sy}(k+1)
    + \nonumber \\
    &+T_j^s \cb{\rho}_j(k+1) + T_j^s \cb{\omega}_j(k). \nonumber
\end{align}
Bearing in mind that the aim is to express $\cb{r}_{i,j}^{sa}(k+1)$ in terms of its previous form $\cb{r}_{i,j}^{sa}(k)$, the forthcoming step is to equalise $\cb{x}_j(k)$ as
\begin{align}\label{eq: equavilence of xjk}
    \cb{x}_j(k) = \cb{y}_{i,j}^s (k) - \cb{\phi}_{i,j}^{sy} (k) - \cb{\rho}_j(k).
\end{align}
Substituting \eqref{eq: equavilence of xjk} into \eqref{eq: Res Calculation2} and considering the dependencies of UIO parameters \eqref{eq: UIOSec_3}-\eqref{eq: UIOSec_5}, we have
\begin{align}
    \cb{r}_{i,j}^{sa}(k+1)  &= F_j^s \cb{r}_{i,j}^{sa}(k) + T_j^s\cb{\phi}_{i,j}^{sy}(k+1) +  \label{eq: Res Calculation3} \\
    & - T_j^sA_{jj}^d\cb{\phi}_{i,j}^{sy}(k) 
    + \cb{\chi}_{i,j}^{s}(k), \nonumber
\end{align}where $\cb{\chi}_{i,j}^{s}(k) = T_j^s\cb{\rho}_j(k+1) - T_j^sA_{jj}^d\cb{\rho}_j(k) + T_j^s\cb{\omega}_j(k)$ collects the noise related term. Equation \eqref{eq: Res Calculation3} implies that there are explicit relations between residuals and SFDIA bias injections, based on which it would be possible to infer the bias vector $\cb{\phi}_{i,j}^s$. Another observation from \eqref{eq: Res Calculation3} is that, in ideal cases, it can determine only one bias entry of $\cb{\phi}_{i,j}^s$ as matrix $T_j^s$ does not possess full-column rank. Thus, extra legitimate information is necessary for the bias injection reconstruction.

\subsection{Physical-Interconnection-Aware Bias Reconstruction}
After detecting the SFDIA \eqref{eq: FDIA against secondary controller} through UIO alarming conditions \eqref{eq: detection conditions under SFDIA} at time instant $k_{i,j}^{sa}$, the bias injected into the neighboring current measurement, i.e., $\phi_{i,j,I}^{sy,re}(k+1), k\ge k_{i,j}^{sa}$, can be reconstructed through the residual $\cb{r}_{i,j}^{sa}$ and observed voltage bias injection $\phi_{i,j,V}^{sy,ob}$
as
    \begin{align}\label{eq: SFDIA bias reconstruction}
        \phi_{i,j,I}^{sy,re} (k+1) = (\tilde{\cb{t}}_{j2}^s)^{\rm T}\Big( T_j^sA_{jj}^d&\cb{\phi}_{i,j}^{sy,re}(k) -\cb{t}_{j1}^s\phi_{i,j,V}^{sy,ob}(k+1) + \nonumber \\
        &
        + \cb{\psi}_{i,j}^{sa}(k)\Big), k\ge k_{i,j}^{sa},
    \end{align}
where $\cb{\phi}_{i,j}^{sy,re}(k) = [\phi_{i,j,V}^{sy,ob}(k),\phi_{i,j,I}^{sy,re}(k)]^{\rm T}$ includes the previous reconstructed current bias injection, $\cb{\psi}_{i,j}^{sa}(k) = \cb{r}_{i,j}^{sa}(k+1)-F_j^s\cb{r}_{i,j}^{sa}(k)$ represents the residual discrepancy term under attacks, $T_j^s = [\cb{t}_{j1}^s, \cb{t}_{j2}^s]$, and $\tilde{\cb{t}}_{j2}^s = \frac{\cb{t}_{j2}^s}{(\cb{t}_{j2}^s)^{\rm T}\cb{t}_{j2}^s}$. 

The voltage injection bias $\phi_{i,j,V}^{sy}$ is hard to be inferred through the information obtained from existing measuring facilities, but can be observed from extra sensor readings based on the physical couplings between DERs. This is idea is made feasible as the electrical and communication graphs share the same topology, and thus the electrical coupling between DERs can be utilised to observe the legitimate voltage information transmitted through the corresponding communication link. After installing current sensors on the power lines interconnecting DERs, from which the power line current is securely measured as $I_{ij}^{se}$, the voltage bias injection can be accurately observed when DER $i$'s local sensor channels are free from attacks, i.e.,
\begin{align}\label{eq: Observed SFDIA voltage bias}
    \phi_{i,j,V}^{sy, ob} (k) &= \cb{\kappa}^{\rm T} \cb{y}_{i,j}^{s}(k) - \underbrace{\big(\cb{\kappa}^{\rm T}\cb{y}_i^p(k) - R_{ij}I_{ij}^{se}(k)\big)}_{V_{i,j}^{ob}(k)} \nonumber \\
    &\approx \phi_{i,j,V}^{sy} (k), k \ge k_{i,j}^{sa},
\end{align}because the inductive part of DC power line is negligible, i.e., $L_{ij} \approx 0$, where $V_{i,j}^{ob}$ is the DER $j$'s PCC voltage observed by DER $i$ through the power line coupling. The observation $\phi_{i,j,V}^{sy, ob} (k)$ can be available at the alarming time instant $k_{i,j}^{sa}$ as only local information is required, but the reconstruction scheme \eqref{eq: SFDIA bias reconstruction} will be activated in the next time instant $k_{i,j}^a+1$ to make the reconstructed bias accurate enough because it needs to incorporate the bias information at the previous time instant. {\color{blue}Benefiting from the development of non-contact measuring technologies like the hall effect current sensor, which operates through generating hall voltage proportional to the current and magnetic field and has been widely adopted in commercial solar inverters \cite{HallSensor1,HallSensor2}, the power line current can be measured without breaking the original circuit connection.}


\subsection{Theoretical Performance Analysis} 

Although the reconstruction scheme \eqref{eq: SFDIA bias reconstruction} is derived from the theoretical links between residuals and bias injections \eqref{eq: Res Calculation3}, the accuracy of reconstructed bias can be still affected by the initial reconstruction error and system noise. It is necessary to theoretically analyse the reconstruction performance when suffering from these unknown disturbances.

\begin{Propos}\label{Pros: Bias Reconstruction under SFDIA}
    Under the bias reconstruction scheme \eqref{eq: SFDIA bias reconstruction}, when the bias observations are accurate as claimed in \eqref{eq: Observed SFDIA voltage bias}, the steady-state current bias reconstruction error, denoted by $\varphi_{i,j,I}^{sy}(k) = \phi_{i,j,I}^{sy}(k) - \phi_{i,j,I}^{sy,re}(k)$, will be bounded by 
    \begin{align}\label{eq: SFDIA bias reconstruction bound}
        |\varphi_{i,j,I}^{sy}(\infty)| \le \frac{(|\tilde{\cb{t}}_{j2}^s|)^{\rm T}\bar{\cb{\chi}}_{i,j}^{s}}{1-\eta_j^s},
    \end{align}
    if the eigenvalue of the reconstruction dynamics \eqref{eq: SFDIA bias reconstruction}, i.e., $\eta_j^s$, satisfies
    \begin{align}\label{eq: SFDIA bias reconstruction stability condition}
        |\eta_j^s| = |(\tilde{\cb{t}}_{j2}^s)^{\rm T}T_j^sA_{jj}^d\cb{\iota}| < 1,
    \end{align}
    where $|\cb{\chi}_{i,j}^s(k)| \le \bar{\cb{\chi}}_{i,j}^{s} = \big(|T_j^s|\bar{\cb{\rho}}_j + |T_j^sA_{jj}^d|\bar{\cb{\rho}}_j + |T_j^s|\bar{\cb{\omega}}_j\big)$ denotes the bound of the noise-related term in \eqref{eq: Res Calculation3}. 
\end{Propos}

The proof of Proposition \ref{Pros: Bias Reconstruction under SFDIA} is conducted in a straightforward manner through calculating the expressions of $\psi_{i,j,I}^{sy}$ considering initial reconstruction errors and system noises, whose details are provided in Appendix \ref{appendix: proof of proposition 1}. The establishment of \eqref{eq: SFDIA bias reconstruction stability condition} is critical in obtaining an accurate enough bias reconstruction result, under which the steady-state bounded reconstruction error \eqref{eq: SFDIA bias reconstruction bound} determined by the system noises' bounds can be guaranteed. {\color{blue}Specifically, the impact of initial error on the steady-state bias reconstruction's accuracy can be eliminated by the stable eigenvalue $\eta_j^s$ as the associated reconstruction error term will decay asymptotically to zero with time. While the system noises can persistently affect the reconstruction accuracy, the accumulated reconstruction error by system noises is proved to be bounded by a known threshold under stable eigenvalue $\eta_j^s$.} In the forthcoming part, we will analytically analyse the establishment of \eqref{eq: SFDIA bias reconstruction stability condition} under different RLC filter parameters by adopting appropriate approximations.

\begin{Propos}\label{Pros: Stability Condition Analyse}
    When the products of sampling time $T_{samp}$ and matrix $A_{jj}$'s entries, denoted by $a_{jj}^{rc}, \forall r, c \in \{1,2\}$, satisfy $|a_{jj}^{rc}T_{samp}| < 1$, then $\eta_j^s$ can be approximated as
    \begin{align}\label{eq: approximation of reconstruction eigenvalue}
        \eta_j^s \approx \eta_j^{s,appr} = 1 + a_{jj}^{22} T_{samp},
    \end{align}under which \eqref{eq: SFDIA bias reconstruction stability condition} will definitely hold since $a_{jj}^{22} < 0$.
\end{Propos}

The proof of Proposition \ref{Pros: Stability Condition Analyse} is proceed with the Taylor expansion of matrix exponential $e^{A_{jj}T_{samp}}$ and the condition $|a_{jj}^{rc}T_{samp}| < 1, \forall r,c \in \{1,2\}$, whose details are elaborated in Appendix \ref{Pros: Proof of Pros 2}. Equation \eqref{eq: approximation of reconstruction eigenvalue} means that the eigenvalue $\eta_j^s$ is mainly determined by $A_{jj}$'s entry $a_{jj}^{22}$ and the sampling time $T_{samp}$ while is weakly associated with $A_{jj}$'s other entries and is irrelevant with the UIO matrix $T_j^s$. Moreover, the assumption of $|a_{jj}^{rc}T_{samp}| < 1$ is reasonable because setting $T_{samp}$ the same as the switching frequency of converters, e.g., 1kHz, can easily make it hold. Since numerous approximations have been adopted in the proof, the accuracy of approximation \eqref{eq: approximation of reconstruction eigenvalue} and the establishment of condition \eqref{eq: SFDIA bias reconstruction stability condition} under various RLC filter parameters are validated through extensive numerical results in Section \ref{section: validation part}.

\subsubsection{Impact Mitigation} The reconstructed bias vector $\cb{\phi}_{i,j}^{sy,re}$ is subtracted from the received compromised data $\cb{y}_{i,j}^s$, i.e.,

\begin{align}\label{eq: SFDIA imapct mitigation}
    \cb{y}_{i,j}^{s,co} (k) = \cb{y}_{i,j}^s (k) - \cb{\phi}_{i,j}^{sy,re}(k), k \ge k_{i,j}^{sa} + 1,
\end{align}which is then used as corrected data to calculate the legitimate secondary control input as in \eqref{eq: Secondary control input}. {\color{blue}Since the reconstructed bias $\cb{\phi}_{i,j}^{sy,re}$ can approach the actual bias $\cb{\phi}_{i,j}^{sy}$ with a bounded error determined the bounds of system noises, the adverse attack impact can almost be eliminated by the mitigation action \eqref{eq: SFDIA imapct mitigation}.}

The implementation details of the proposed detection-triggered impact mitigation scheme is provided in Algorithm \ref{Alg: detection-triggered impact mitigation}. In steps 1-8, the detection residual $\cb{r}_{i,j}^s$ and residual bound $\cb{r}_{i,j}$ are calculated to check the establishment of conditions \eqref{eq: detection conditions under SFDIA}. {\color{blue}The alarm signal $\varrho_{i,j}^s$ will be flagged to trigger the bias reconstruction phase once any anomaly is perceived. Steps 9-20 elaborate the bias reconstruction process and impact mitigation action. If $\varrho_{i,j}^s$ is flagged and there exists current sensor on power line $(i,j)$, the power line current will first be utilised to observe the voltage bias injection $\phi_{i,j,V}^{sy,ob}$ through \eqref{eq: Observed SFDIA voltage bias}. Then, the current bias injection $\phi_{i,j,I}^{sy,re}$ can be accurately reconstructed from the residuals under attacks $\cb{r}_{i,j}^{sa}$ and $\phi_{i,j,V}^{sy,ob}$ according to \eqref{eq: SFDIA bias reconstruction}. When there exits no current sensor within DER $i$ for power line $(i,j)$, then the reconstructed bias will be set as $\cb{\phi}_{i,j}^{sy,re} = \cb{y}_{i,j}^s - \cb{y}_i$ such that the data received from DER $j$ is unused.}3 Finally, $\phi_{i,j,I}^{sy,re}$ will be subtracted from the compromised data $\cb{y}_{i,j}^s$ to eliminate the attack impacts on the secondary controller. {\color{blue}Under the activated mitigation action, the safety criteria that implies the completion of attack impact elimination is
\begin{align}\label{eq: attack impact elimination}
    \tilde{\alpha}_i(k) = \sum_{j\in\mathcal{N}_i^c}\Big|\frac{\cb{y}_{i,j}^{s,co}(k)}{I_{tj}^s} - \frac{\cb{y}_{i}^p(k)}{I_{ti}^s}\Big| \le \tau,
\end{align}where threshold $\tau>0$ is used to judge the accomplishment of load sharing from DER $i$'s view. The satisfaction of \eqref{eq: attack impact elimination} does not mean the mitigation actions can be withdrawn as the SFDIA will exist all the time, and thus appropriate cyber recovery schedule should be made to remove the intruder from the communication network \cite{liu2023cyber}, which still requires substantial efforts and are as our future works.}

{\footnotesize
\begin{algorithm}
\captionsetup{font={small}}
\caption{Detection-Triggered Impact Mitigation under $\cb{\phi}_{i,j}^{sy}$}\label{Alg: detection-triggered impact mitigation}
\begin{algorithmic}[1]
\Require DER parameters $A_{jj}^d, \cb{m}_j^d$, and $\cb{b}_j^d$; UIO$_{i,j}^s$ parameters $F_j^s, T_j^s, H_j^s, \hat{K}_j^s$; Communicated DER $j$'s output $\cb{y}_{i,j}^s$ and converter command $u_{i,j}^s$; Residual bound parameters $\nu_j^s, \varsigma_j^s$; Secure power line current measurements $I_{ij}^{se}$.
\Ensure Detection residual $\cb{r}_{i,j}^s$ and corresponding bound $\bar{\cb{r}}_{i,j}^s$; Attack alarm flag $\varrho_{i,j}^s$; Reconstructed current bias injection $\phi_{i,j,I}^{sy,re}$.
\State \textit{\textbf{Detection Phase}}
\State Calculate the detection residual $\cb{r}_{i,j}^s$ according to \eqref{eq: UIO for secondary controller}-\eqref{eq: Residual calculation}
\State Calculate the residual bound $\bar{\cb{r}}_{i,j}^s$ according to \eqref{eq: primary residual upper bound}
\If{Either of \eqref{eq: detection conditions under SFDIA} is satisfied}
    \State Flag the alarm signal $\varrho_{i,j}^s = 1$
\Else
    \State Reset the alarm signal $\varrho_{i,j}^s = 0$
\EndIf
\State \textit{\textbf{Bias Reconstruction and Impact Mitigation Phase}}
\If{The alarm signal $\varrho_{i,j}^s$ is flagged}
\If{Current sensor has been deployed in DER $i$ for power line $(i,j)$}
\State Observe the voltage bias injection through \eqref{eq: Observed SFDIA voltage bias}
\State Reconstruct the current bias injection through \eqref{eq: SFDIA bias reconstruction}
\Else
\State {\color{blue}Set $\cb{\phi}_{i,j}^{sy,re} = \cb{y}_{i,j}^s - \cb{y}_i$; \Comment{Discard the data received from DER $j$}
}
\EndIf
\Else
\State Reset the current bias injection as zero
\EndIf
\State Eliminate the SFDIA's impacts on secondary controllers through \eqref{eq: SFDIA imapct mitigation}
\end{algorithmic}
\end{algorithm}}
{\color{blue}
\section{Cost-Effective Current Sensor Deployment}
Given the high budget of deploying local current sensors on all power lines, a cost-effective deployment strategy is necessary to facilitate the proposed method's practical application. This section will first introduce a useful method that utilises the output current $I_{ti}$ and PCC voltage $V_{i}$ to estimate the sum of power line currents outputted from DER $i$, denoted by $I_{i,se}^{sum}=\sum_{j\in\mathcal{N}_i^{el}}I_{ij}^{se}$, which can help avoid the deployment of redundant current sensors. Based on this, a cost-effective sensor deployment scheme will be then presented to keep only the necessary current sensors required by the secondary control performance.

\subsection{Estimation of Power Line Current Sum}
According to the circuit of DER $i$, the relationships between the original measurements for primary control, i.e., $I_{ti}, V_i$, and $I_{i,se}^{sum}$ can be written as
{\small\begin{align}\label{eq: estimated sum of current loads}
    I_{i,se}^{sum}(k) = I_{ti}(k) &- \frac{C_{ti}}{T_{samp}}\big(V_i(k) - V_i(k-1)\big) - \hat{I}_{Li}^{sum}(k),
\end{align}}where $\hat{I}_{Li}^{sum}$ denotes the sum estimation of currents flowing over the ZIP load obtained either from historical load data \cite{lai2021sizing,vossos2014energy} or via load forecasting methods \cite{trivedi2022data,sun2019using}. In the ideal case, the estimated sum of load currents should satisfy $\hat{I}_{Li}^{sum} = I_{Li} + \frac{V_{i}}{Z_{Li}}$. Although there can be some errors between the sum estimation and actual one due to the elusive load usage behaviours, the error can be kept below a certain threshold from the long term such that the resulted mitigation actions would still be able to significantly decrease the attack impact as validated in Section \ref{section: validation part}-B.

The utilisation of \eqref{eq: estimated sum of current loads} can decrease the number of extra current sensors required by the proposed mitigation scheme. In particular, since the sum of currents outputted from DER $i$ is known, it is not necessary to measure all power line currents and one redundant current sensor can be removed. For example, if DER $i$ has only one power line connecting it to other DERs, then the estimated current sum can be directly regarded as its power line current. 

\subsection{Cost-Effective Deployment Strategy of Current Sensors}
Besides integrating the estimated sum of power line currents \eqref{eq: estimated sum of current loads}, the deep analysis of secondary control's objective can further decrease the number of deployed current sensors. The secondary controller \eqref{eq: Secondary control input} is designed to achieve load sharing among DERs through an essential consensus principle, and one of its sufficient condition is to have a connected communication graph \cite{tucci2018stable}. However, in practice, more dense communication network may be developed to guarantee the efficient and robust collaboration of DERs in microgrids. The dense communication network means that the corresponding graph would be still connected even though one or more edges are removed, under which load sharing can be still achieved, but in a slower convergence rate. Therefore, from a cost-effective perspective, the extra current sensors can be strategically deployed on the power lines that makes the “secured” communication links form a spanning tree, i.e., a connected sub-graph possessing the least number of edges. Here, the “secured” communication link means that the legitimate data transmitted over it can be reconstructed by the proposed scheme. A cost-effective deployment strategy of current sensors will be presented in the subsequent part.

The deployment of current sensors consists of two phases as shown in Algorithm \ref{Alg: current sensor deployment}. \textit{Phase I}: Selecting a spanning tree sub-graph $\mathcal{G}_c'=\{\mathcal{A},\mathcal{E}_c'\}$ of $\mathcal{G}_c$ that connects DER nodes with the least number of edges and forms no circle. The selection is started from removing edges from the original edge set $\mathcal{E}_c$, where the removed edges and associated nodes are collected in $\mathcal{E}_c^{rm}$ and $\mathcal{A}^{rm}$, respectively. The removed edge should be a part of a cycle that is found though depth-first-search based function, i.e., \textit{findcycle}, which has linear time and space complexity with respect to the size of searched graph. In step 3, the edge set returned by \textit{findcycle} is recorded as $\mathcal{E}_c^{cy}$. In steps 5-11, for each edge $(i,j)$ in $\mathcal{E}_c^{cy}$, it will be used to update $\mathcal{E}_c^{rm}$ and $\mathcal{A}^{rm}$ if nodes $i$ or $j$ has been covered by the removed node sets. Otherwise, through steps 12-16, the edge of $\mathcal{E}_c^{cy}$ covering the highest connectivity node will be chosen to update $\mathcal{E}_c^{rm}$ and $\mathcal{A}^{rm}$. Then, the spanning tree sub-graph can be obtained as $\{\mathcal{A}, \mathcal{E}_c/\mathcal{E}_c^{rm}\}$ as in step 17. The reason behind removing the edge within cycle $\mathcal{E}_c^{cy}$ that has overlap with the removed node set $\mathcal{A}^{rm}$ is to keep the size of $\mathcal{A}^{rm}$ as small as possible. Since the electrical and communication graphs have the same topology, the resulted spanning tree edge set $\mathcal{E}_c/\mathcal{E}_c^{rm}$ can be mapped into the electrical graph equally. Therefore, the smaller $\mathcal{A}^{rm}$ means that the mapped spanning tree sub-graph can cover more DER nodes that have complete 
adjacent physical couplings, under which more redundant current sensors can be removed based on the estimated sum of power line currents \eqref{eq: estimated sum of current loads} in the next phase. Moreover, when there exists no overlap between $\mathcal{E}_c^{cy}$ and $\mathcal{A}^{rm}$, the edge covering the highest connectivity node will be used to update the removed edge and node sets, which can maximise the overlap possibility in the subsequent cycle traversal.


\textit{Phase II}: Deploying current sensors on power lines based on the obtained spanning tree edge set $\mathcal{E}_c/\mathcal{E}_c^{rm}$ and \eqref{eq: estimated sum of current loads}. In step 21, the sensor deployment positions, which are recorded as directed edges as the current sensor is deployed within local DER, are initialised as $\mathcal{E}_{el}'$ that contains all spanning tree edges in $\mathcal{E}_c/\mathcal{E}_c^{rm}$. Through steps 22-26, if all adjacent physical couplings of DER $i$ are included in $\mathcal{E}_{el}'$, then there is one redundant sensor deployment according to \eqref{eq: estimated sum of current loads} and one arbitrary adjacent edge can be removed. In general, the number of deployed current sensors is calculated as $|\mathcal{A}| + |\mathcal{A}^{rm}| - 2$, which is proportional to the network size and density since $|\mathcal{A}^{rm}|$ will increase as the number of included circles.

\begin{figure*}[tb]
    \centering
    \includegraphics[width = 18cm]{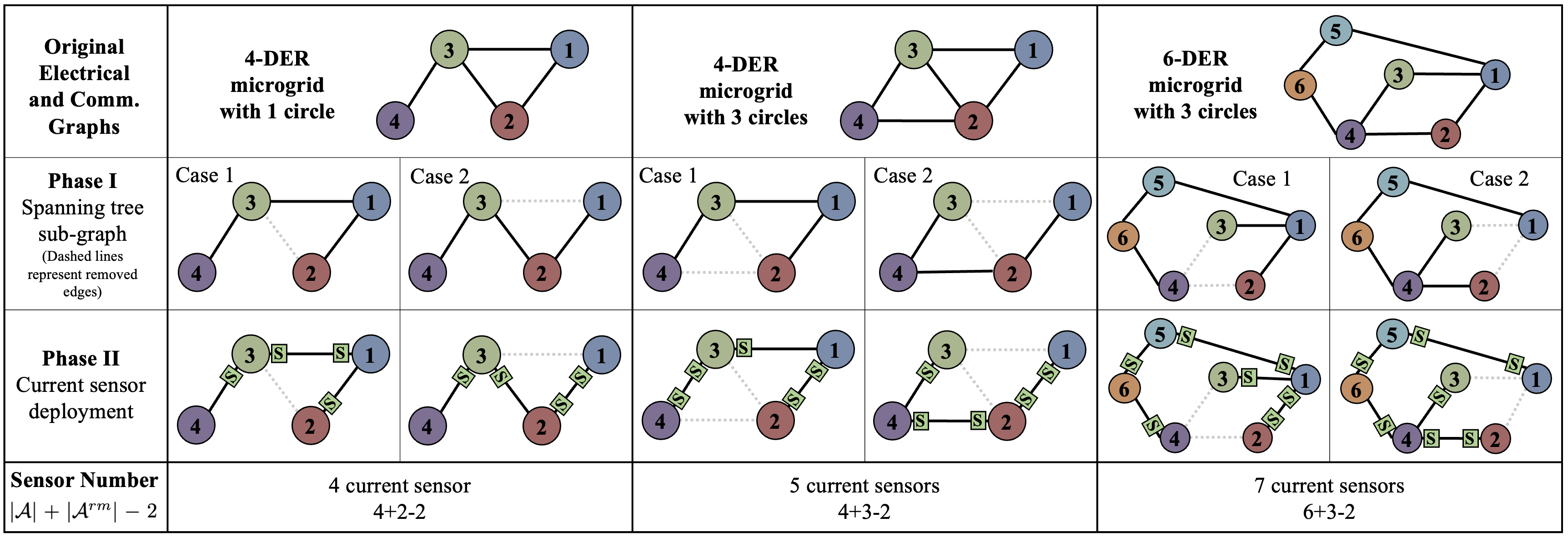}
    \captionsetup{font={footnotesize}}
    \caption{\color{blue}This figure demonstrates the deployment of current sensors on power lines under electrical and communication graphs with various size and density. It is clear that as the increase of network size and density, the number of deployed sensors will grow accordingly.}
    \label{fig:CurrentSensorDeploymentDemonstration}
\end{figure*}

Demonstrative examples are given in Fig. \ref{fig:CurrentSensorDeploymentDemonstration} to illustrate the results of Algorithm \ref{Alg: current sensor deployment}. For a 4-DER microgird with 1 circle, 4 current sensors are needed to ensure load sharing under worst-case SFDIAs. When the number of circles grows to 3, one more current sensor is needed to maintain the control performance. Under the same number of circles, if the number of nodes increases to $6$, two more current sensors are required to cover the extended network size. Therefore, when the electrical and communication networks' size and density grow, the required security investment will also increase with a linear rate. Moreover, it is noted that Algorithm \ref{Alg: current sensor deployment} may output multiple sensor deployment solutions, implying that other considerations like the criticality of communication link and DER node can be incorporated to further optimise the result. Future efforts are still needed to conduct insightful analysis towards this direction.}

{\footnotesize
\begin{algorithm}
\captionsetup{font={small}}
\caption{\color{blue}Cost-Effective Current Sensor Deployment}\label{Alg: current sensor deployment}
\begin{algorithmic}[1]
\Require Electrical graph $\mathcal{G}_{el}$, Communication graph $\mathcal{G}_c$
\Ensure Spanning tree edge set $\mathcal{E}_c/\mathcal{E}_c^{rm}$, Current sensor positions $\mathcal{E}_{el}'$
\State \textbf{Phase I: Select Spanning Tree Edge Set}
\State Set empty removed edge and node sets $\mathcal{E}_c^{rm}, \mathcal{A}^{rm}$;
\State $\mathcal{E}_c^{cy}=$\ \textit{findcycle}($\{\mathcal{A},\mathcal{E}_c\}$);
\While{$\mathcal{E}_c^{cy}$ is not empty}
\For{Edge $(i,j)\in\mathcal{E}_c^{cy}$}
\If{$i\in\mathcal{A}^{rm}$ or $j\in\mathcal{A}^{rm}$}
\State Update removed edge set $\mathcal{E}_c^{rm} = \mathcal{E}_c^{rm} \cup (i,j)$;
\State Update removed node set $\mathcal{A}^{rm} = \mathcal{A}^{rm} \cup \{i,j\}$;
\State Break;
\EndIf
\EndFor
\If{There is no overlap between $\mathcal{E}_c^{cy}$ and $\mathcal{A}^{rm}$}
\State Choose $(i,j) \in \mathcal{E}_c^{cy}$ that covers the highest connectivity node;
\State Update removed edge set $\mathcal{E}_c^{rm} = \mathcal{E}_c^{rm} \cup (i,j)$;
\State Update removed node set $\mathcal{A}^{rm} = \mathcal{A}^{rm} \cup \{i,j\}$;
\EndIf
\State $\mathcal{E}_c^{cy}=$\ \textit{findcycle}($\{\mathcal{A},\mathcal{E}_c/\mathcal{E}_c^{rm}\}$);
\EndWhile
\State Obtain the spanning tree sub-graph $\{\mathcal{A}, \mathcal{E}_c/\mathcal{E}_c^{rm}\}$;
\State \textbf{Phase II: Deploy Current Sensors on Power Lines}
\State Let $\mathcal{E}_{el}'$ cover all spanning tree edge set $\mathcal{E}_c/\mathcal{E}_c^{rm}$;
\For{$i \in \mathcal{A}$}
\If{All adjutant edges of node $i$ are included in $\mathcal{E}_{el}'$}
\State Remove one arbitrary adjacent edge from $\mathcal{E}_{el}'$;
\EndIf
\EndFor
\end{algorithmic}
\end{algorithm}}

Compared with the hidden secure network \cite{zuo2020resilient} and digital twin \cite{saad2020implementation}, the proposed bias reconstruction scheme \eqref{eq: SFDIA bias reconstruction} does not need to run the extra hardware along with the original system. {\color{blue}Moreover, the cost of installing current sensors, whose number grows linearly with respect to the network size and density, would be much cheaper than developing a standalone communication network or digital system.} While the detection-triggered bias reconstruction scheme \eqref{eq: SFDIA bias reconstruction} can also avoid the limitation of requiring at least one trustworthy communication link between DERs to guarantee the resilient control performance as in \cite{sahoo2020multilayer}. Hence, the primary difference of the proposed detection-triggered bias reconstruction and impact mitigation scheme from existing literature is that it can effectively improve the impact mitigation performance in a cost-efficient manner.

{\color{blue}
It has great importance to discuss the adaptability of the proposed mitigation method to other types of FDIAs besides \ding{172} as demonstrated by Fig. \ref{fig:AttackTypeIllustration}. Since this paper focuses on reconstructing the bias injections on the interaction links between DERs, a natural consideration is to apply it the cases with corrupted local measurements. When the local measurements sent to secondary controller are compromised under SFDIA \ding{174}, the proposed bias reconstruction scheme can be seamlessly used to reconstruct the legacy current measurement once one of its neighboring DERs can accurately estimate its PCC voltage through the physical interconnection. For the scenario where the local measurements received by primary controller are corrupted by PFDIA \ding{176}, the voltage estimation transmitted through the secondary communication network may not satisfy the primary control's real-time requirement. The proposed reconstruction scheme will be thus not able to reconstruct the legacy current and voltage measurements from the residual of one UIO due to the insufficient observability. Further improvements are required to make the bias injections on both current and voltage measurements observable from the UIO residuals. Moreover, the investigation of mitigation strategies under corrupted control commands, i.e., FDIAs \ding{173} and \ding{175}, has been well addressed by existing literature \cite{zuo2020distributed,9039721,9184978,9600614}. But they all require the sensor channels to be free from attacks, and thus the reconstructed legal measurement from the proposed method can be a useful complement for them when both actuator and sensor channels are subject to bias injections. }

\section{Simulation Studies and Experimental Validations}\label{section: validation part}
{\color{blue}This section first verifies the reconstruction stability under varying electrical parameters, and then validates the proposed detection-triggered impact scheme's effectivess in Matlab/Simulink and HIL testbeds. Specifically, a 16-DER microrgid is established in Matlab/Simulink to demonstrate the proposed method's applicability to large-scale systems considering load variations, DERs plugging-in, and security budget limitation. By integrating Typhoon HIL, raspberry pis, and EXata network modelling software, a HIL 6-DER microgrid is setup to test the proposed method's performance in the presence of practical communication protocols and realistic attack events. 
}




\begin{figure*}[]
  \centering
  \includegraphics[width=18cm]{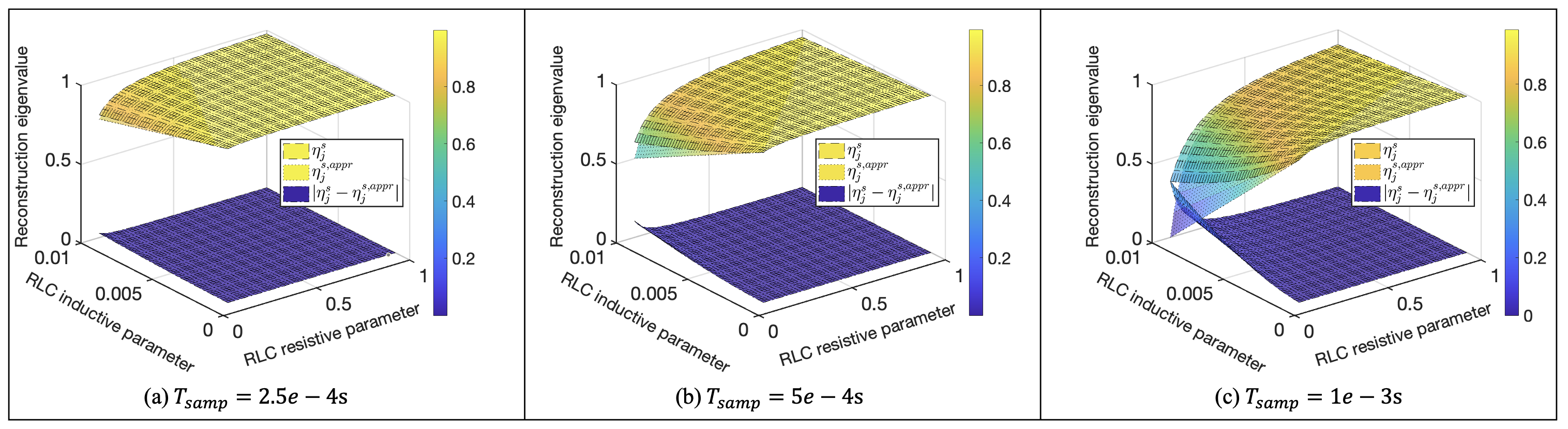}
  \captionsetup{font={footnotesize}}
  \caption{This figure verifies that $\eta_j^s<1$ is always stable under varying $R_{ti}, L_{ti}$, and $T_{samp}$, and that the approximation $\eta_j^{s,appr}$ is accurate at most of the time except when $\eta_j^s$ is far away from $1$.}\label{fig:ReconstructiveEigenvalueRLCParametersImpact}
\end{figure*}

\subsection{Reconstruction Stability under Varying Parameters}
As claimed in Propositions \ref{Pros: Bias Reconstruction under SFDIA}-\ref{Pros: Stability Condition Analyse}, the stability of bias reconstruction scheme \eqref{eq: SFDIA bias reconstruction} is determined by the dynamics' eigenvalue $\eta_j^s$, which can be approximated as $\eta_j^{s,appr}$ that is mainly related to the RLC resistive ($R_{ti}$) and inductive ($L_{ti}$) parameters as well as sampling time $T_{samp}$. Inspired by this, we first show that $|\eta_j^s|$ under varying RLC capacitive parameter $C_{ti}$ can be regarded as constant and is less than $1$. Then, the variation of $|\eta_j^s|<1$ under varying $R_{ti}, L_{ti}$, and $T_{samp}$ is demonstrated and explained with the approximation \eqref{eq: approximation of reconstruction eigenvalue} from Proposition \ref{Pros: Stability Condition Analyse}. 

\subsubsection{Constant and Stable $\eta_j^s$ under Varying $C_{ti}$}
\begin{figure}
  \centering
  \includegraphics[width=5cm]{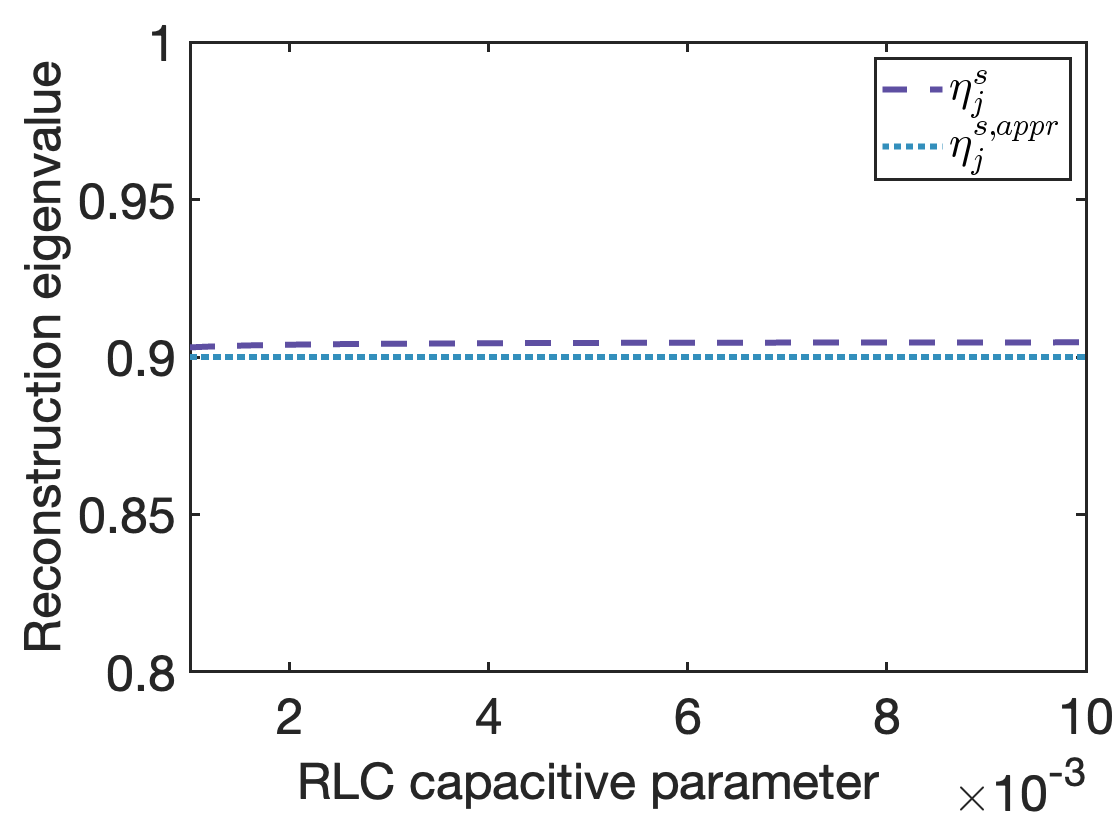}
  \captionsetup{font={footnotesize}}
  \caption{This figure shows that the $\eta_j^s$ under varying $C_{ti}$ is stable and can be regarded as constant, under which the approximation error $|\eta_j^s - \eta_j^{s,appr}|$ is negligible.}\label{fig:ReconstructiveEigenvalueCapactiveImpact}
\end{figure}

: By keeping $R_{ti}, L_{ti}$, and $T_{samp}$ invariant as $0.2, 1e-3$, and $5e-4$, respectively, and perturbing $C_{ti}$ within $[1e-3, 10e-3]$F, the values of $\eta_j^s$ and $\eta_j^{s,appr}$ are shown in Fig. \ref{fig:ReconstructiveEigenvalueCapactiveImpact}. Clearly, $\eta_j^s$ changes slowly around $0.9<1$ and the variation range is smaller than $2e-3$. Moreover, the corresponding approximation error $|\eta_j^s - \eta_j^{s,appr}|$ is bounded by $5e-3$, indicating that the approximation \eqref{eq: approximation of reconstruction eigenvalue} is accurate enough.

\subsubsection{Stable $\eta_j^s$ under Varying $R_{ti}, L_{ti}$, and $T_{samp}$} 

The results of $\eta_j^s$ and $\eta_j^{s,appr}$ when varying $R_{ti}$ and $L_{ti}$ within $[0.1,1]\Omega$ and $[1e-3, 10e-3]$H, choosing $T_{samp}$ from $\{2.5e-4, 5e-4, 1e-3\}$s, and keeping $C_{ti}$ as $2.2e-3$ are pictured in Fig. \ref{fig:ReconstructiveEigenvalueRLCParametersImpact}. The first observation is that $|\eta_j^s|$ is always smaller than $1$ regardless of the changes of $R_{ti}, L_{ti}$, and $T_{samp}$, indicating that the reconstruction scheme \eqref{eq: SFDIA bias reconstruction} is always stable under initial errors and system noises. Moreover, the approximation error $|\eta_j^s - \eta_j^{s,appr}|$ is positively associated with $L_{ti}$ and $T_{samp}$ while is negatively related to $R_{ti}$. This phenomenon matches the description of Proposition \ref{Pros: Stability Condition Analyse}, where the approximation can be more accurate with smaller $|a_{jj}^{rc}T_{samp}|$ that can be caused by the decrease of $L_{ti}, T_{samp}$ and increase of $R_{ti}$. The unacceptable approximation error usually occurs with large $L_{ti}$ and small $R_{ti}$, under which the $|\eta_j^s|$ will be far smaller than $1$. Hence, the stability of the reconstruction scheme \eqref{eq: SFDIA bias reconstruction} is verified through theoretical analysis and numerical results.

\begin{figure}[]
  \centering
  \includegraphics[width=9cm]{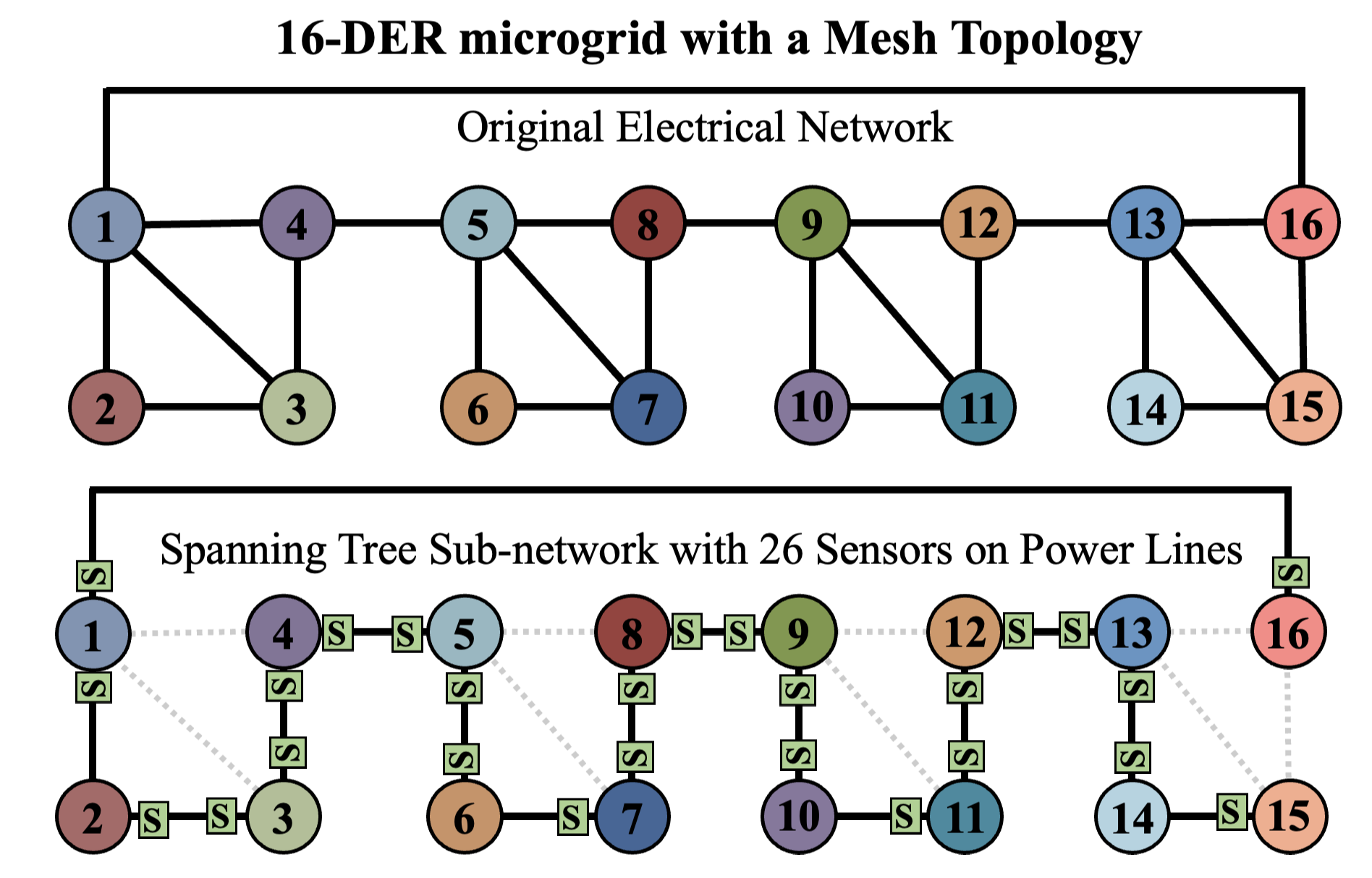}
  \captionsetup{font={footnotesize}}
  \caption{\color{blue}This figure shows the original electrical network of the 16-DER microgrid the secured spanning tree sub-network after deploying 26 sensors on power lines, where dashed lines represent the removed edges that do not require to deploy sensors.}\label{fig:16_DER_microgrid}
  \vspace{-10pt}
\end{figure}

{\color{blue}\subsection{Applicability to the 16-DER Microgrid in Matlab/Simulink}
This subsection demonstrates the reconstruction accuracy and mitigation performance when all communication links are affected by SFDIAs \eqref{eq: FDIA against secondary controller} in the 16-DER microgrid with a mesh topology as shown in Fig. \ref{fig:16_DER_microgrid}. The RLC filter parameters are set as $R_{ti} = 0.2\Omega, L_{ti} = 1e-3$H, and $C_{ti} = 0.5e-3$F, the equivalent resistive loads are $Z_{Li} = 10\Omega$, $\forall i \in \{1,\cdots,16\}$, and the power lines' impedance are $R_{ij} = 1.5\Omega, L_{ij} = 1.8e-6$H, $\forall (i,j) \in \mathcal{E}_{el}$. The reference PCC voltages of DERs are set around $40$V such that their average is equal to it. The communication network topology is considered to be the same as the electrical network topology. The sampling time is $T_{samp}=1e-3$s, under which the stability condition \eqref{eq: SFDIA bias reconstruction stability condition} is satisfied. The event timeline is summarised as follows: 
\begin{itemize}
    \item[a)] The primary controllers are activated at $t=0$s, where DERs 10 and 14 are not physically connected to other DERs.
    \item[b)] The secondary controllers are started at $t=1$s, after which the UIO-based detectors are activated at $t=1.5$s.
    \item[c)] DERs 10 and 14 are plugged into the microgrid to obtain the topology as shown in Fig. \ref{fig:16_DER_microgrid} at $t=3$s.
    \item[d)] Variations occur on the equivalent current loads of DERs 1, 2, 5, 6, 9, 10, 13, 14 at $t=5$s.
    \item[e)] SFDIAs that have continuous and discontinuous bias injections in the form of sine and triangle signals against all communication links are launched at $t=7$s.
\end{itemize}

\begin{figure*}[]
  \centering
  \includegraphics[width=18cm]{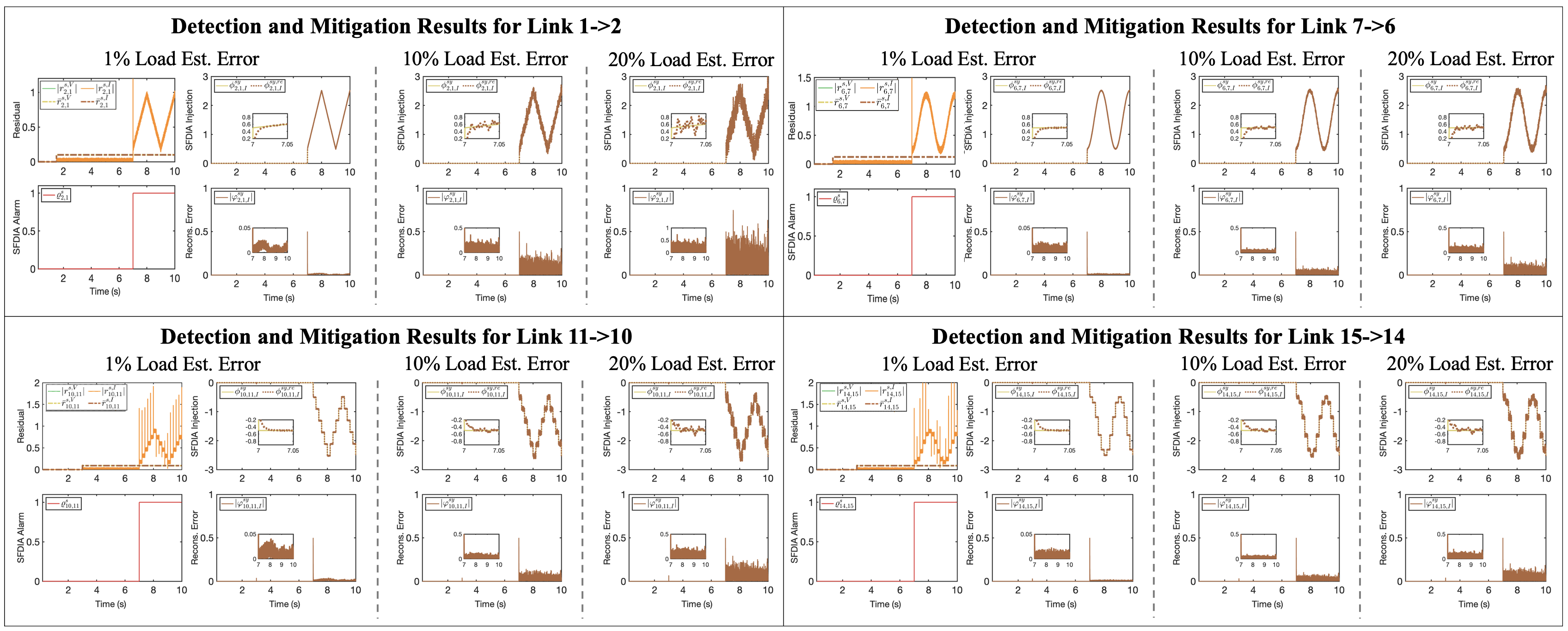}
  \captionsetup{font={footnotesize}}
  \caption{\color{blue}This figure shows the detection and bias reconstruction results for the SFDIAs on links $(2,1), (6,7), (10,11), (14,15)$, which includes both continuous and discontinuous bias injections in the forms of sine and triangle signals. Moreover, the demonstrated reconstruction processes all rely on the estimated load profile since there are no sensor deployed on the corresponding power lines as shown in Fig. \ref{fig:16_DER_microgrid}. Hence, the impacts of different levels of load estimation errors on the reconstruction error are also showcased.}\label{DetectionMitigartionResults_ThreeCases}
  \vspace{-10pt}
\end{figure*}

\begin{figure*}[]
  \centering
  \includegraphics[width=18cm]{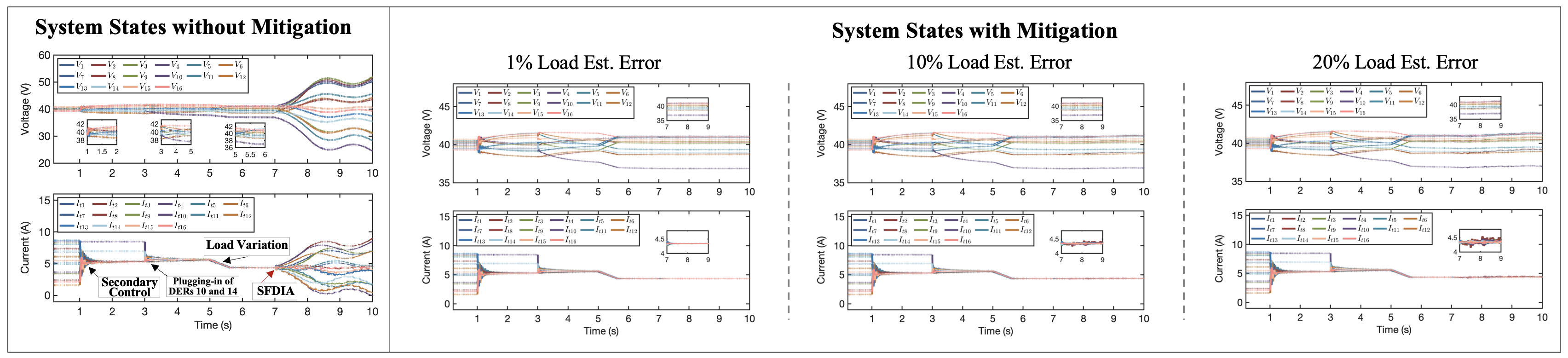}
  \captionsetup{font={footnotesize}}
  \caption{\color{blue}This figure shows the voltage and current states of the 16-DER microgrid without and with impact mitigation strategies when all communication links are subject to SFDIAs. Moreover, the impact mitigation performance under different levels of load estimation errors are also validated.}\label{fig:SystemStates_ThreeCases}
  \vspace{-10pt}
\end{figure*}

\begin{figure}[!h]
  \centering
  \includegraphics[width=8cm]{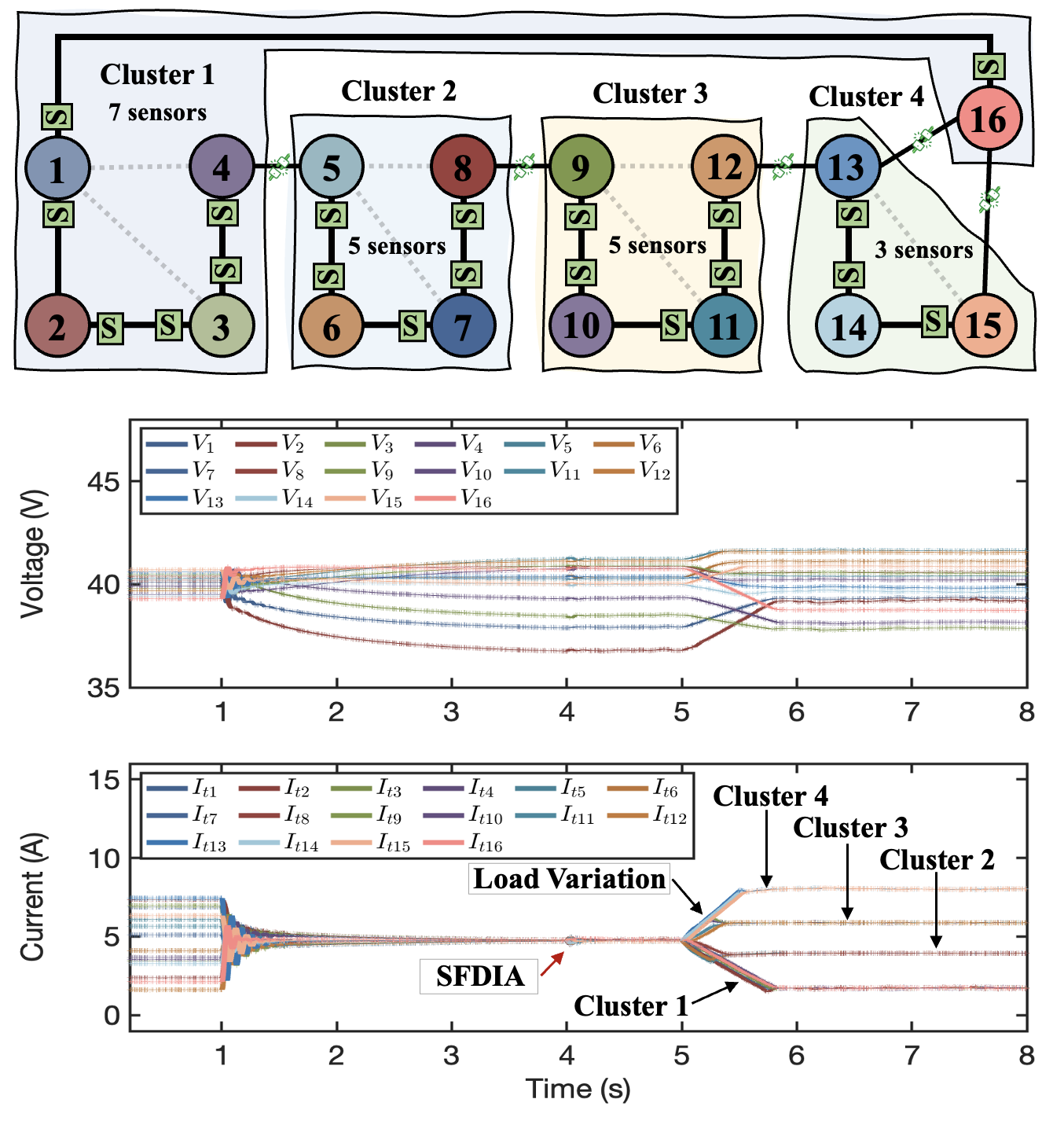}
  \captionsetup{font={footnotesize}}
  \caption{\color{blue}This figure showcases a feasible partition scheme for the 16-DER microgrid to achieve the load sharing objective within partitioned clusters after proactively cutting off power lines $(4,5), (8,9), (12,13), (13,16), (15,16)$ when the number of current sensors is extremely limited.}\label{fig:ExtremeCaseLimitedSensors_DividedClusters}
  \vspace{-10pt}
\end{figure}

\subsubsection{Bias Reconstruction Accuracy}
The detection-triggered bias reconstruction results are shown in Fig. \ref{DetectionMitigartionResults_ThreeCases}. We choose to demonstrate the bias reconstruction accuracy for the SFDIAs on communication links $(2,1), (6,7), (10,11), (14,15)$ since they include the four types of bias injections and their reconstruction processes need to utilise the estimated sum of power line currents \eqref{eq: estimated sum of current loads}. According to the pictured results, it is clear that daily operations before $t=7$s including load variation and DERs plugging-in will not falsely flag the attack alarm and trigger the bias reconstruction scheme. When the SFDIAs are launched at $t=7$s, the bias reconstruction scheme can be immediately triggered to accurately reconstruct both continuous and discontinuous bias injections with the steady-state reconstruction error smaller than $0.05$ under $1\%$ load estimation error. Due to the injected bias's sudden change at $t=7$s, there will be a non-trivial initial reconstruction error that can reach $0.5$. Under stable eigenvalue $\eta_j^s$, the initial reconstruction error will decay asymptotically to zero as time and eventually the reconstruction error can be bounded by a threshold determined by system noises and load estimation errors. For the discontinuous bias injections on links $(10,11), (14,15)$, the detection residual can be occasionally lower than the threshold when the bias injections are small, which can result in unexpected missed alarms. To avoid this issue, the flagged attack alarm will be kept until the received data is validated as normal for $N_{att} = 10$ sampling times.

When the load estimation error is increased from $1\%$ to $10\%$ and $20\%$, respectively, the bias reconstructed accuracy will be affected accordingly. The intuitive observation is the larger load estimation error can result in higher bias reconstruction error. Moreover, the same level of load estimation error may affect the reconstruction error differently when the DER's physical coupling varies. For example, the $10\%$ load estimation error in DER 2 can increase the reconstruction close to $0.5$, while the same level of load estimation error in DERs 6, 10, 14 will only make the reconstruction error close to $0.2$. It can be observed from Fig. \ref{fig:16_DER_microgrid} that DER 1 has 4 physical neighbors while DERs 6, 10, 14 only have 2 physical neighbors. It is noted that the load estimation error's impact on reconstruction error is very similar to that of system noise, and therefore deeper theoretical analysis can be conducted following the procedure in Proposition 1, which are left our future works.

\subsubsection{Mitigation Performance}
The impact mitigation performance of utilising the reconstructed biases from the previous part under different levels of load estimation errors are demonstrated in Fig. \ref{fig:SystemStates_ThreeCases}. In the absence of mitigation strategies, the voltage and current states will be significantly affected by the injected SFDIA biases, under which the load sharing objective can be never achieved. To secure a spanning tree sub-network of the original communication network, 26 current sensors should be deployed on power lines and Fig. \ref{fig:16_DER_microgrid} shows one feasible solution. According to Algorithm \ref{Alg: detection-triggered impact mitigation}, if the detector perceives anomaly on the communication link and the corresponding power line's current cannot be sensed or estimated, then the communication link will be switched off. Otherwise, the bias reconstruction and impact mitigation scheme will be triggered. The results indicate that the reconstructed biases can effectively eliminate the attack impacts and reestablish the load sharing objective under $1\%$ load estimation error. As the increase of load estimation error from $1\%$ to $10\%, 20\%$, there will appear obvious fluctuations on voltages and currents. For example, the steady-state currents will deviate from the expected sharing value with the error bounded by $0.5$A. Although the mitigation performance can be slightly affected by the load estimation error, it is still able to significantly decrease the attack impact from $5$A load sharing error to $0.5$A load sharing error.

When the number of current sensors is too limited to form a spanning tree sub-graph for the 16-DER microgrid, the original microgrid can be partitioned into smaller clusters such that load sharing can be still achieved within these clusters under the worst-case SFDIA case. As shown in Fig. \ref{fig:ExtremeCaseLimitedSensors_DividedClusters}, if the 16-DER microgrid is partitioned into 4 clusters, then the number of sensors required for securing each cluster would only be 7, 5, 5, 3. After detecting SFDIAs, the power lines without current sensors including $(4,5), (8,9), (12,13), (13,16), (15,16)$ will be proactively cut off. The mitigation results indicate that load sharing objective can be successfully achieved within clusters based on the proposed method. The system operator needs to design appropriate partition mechanisms for microgrids to achieve the maximum social profit subject to the security budget for current sensors, which still requires future efforts and is as one of our future works.}

\begin{figure}[]
  \centering
  \subfigure[Overview of microgrid testbed]{\includegraphics[width=8cm]{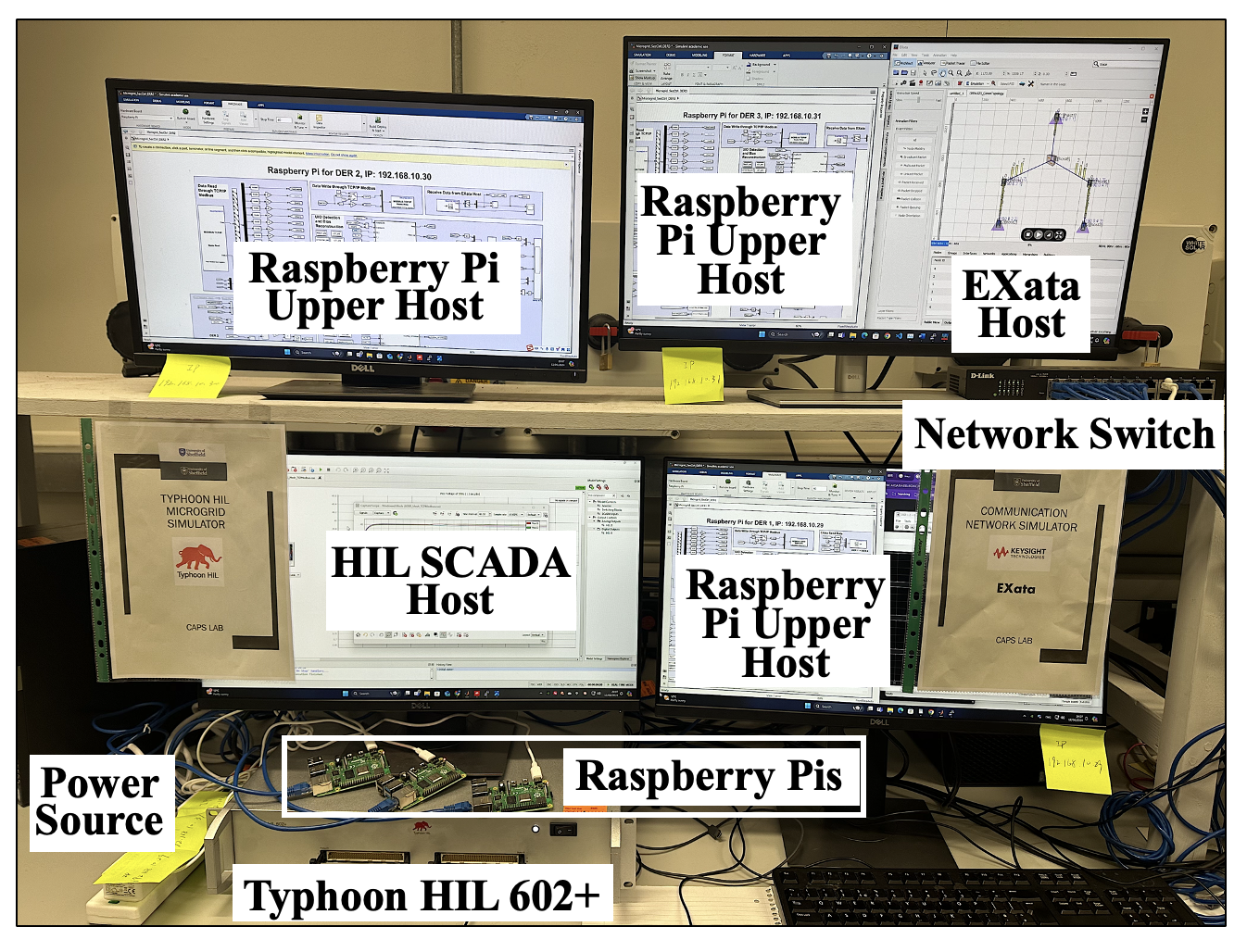}}
  \subfigure[Illustration of component connection]{\includegraphics[width=8cm]{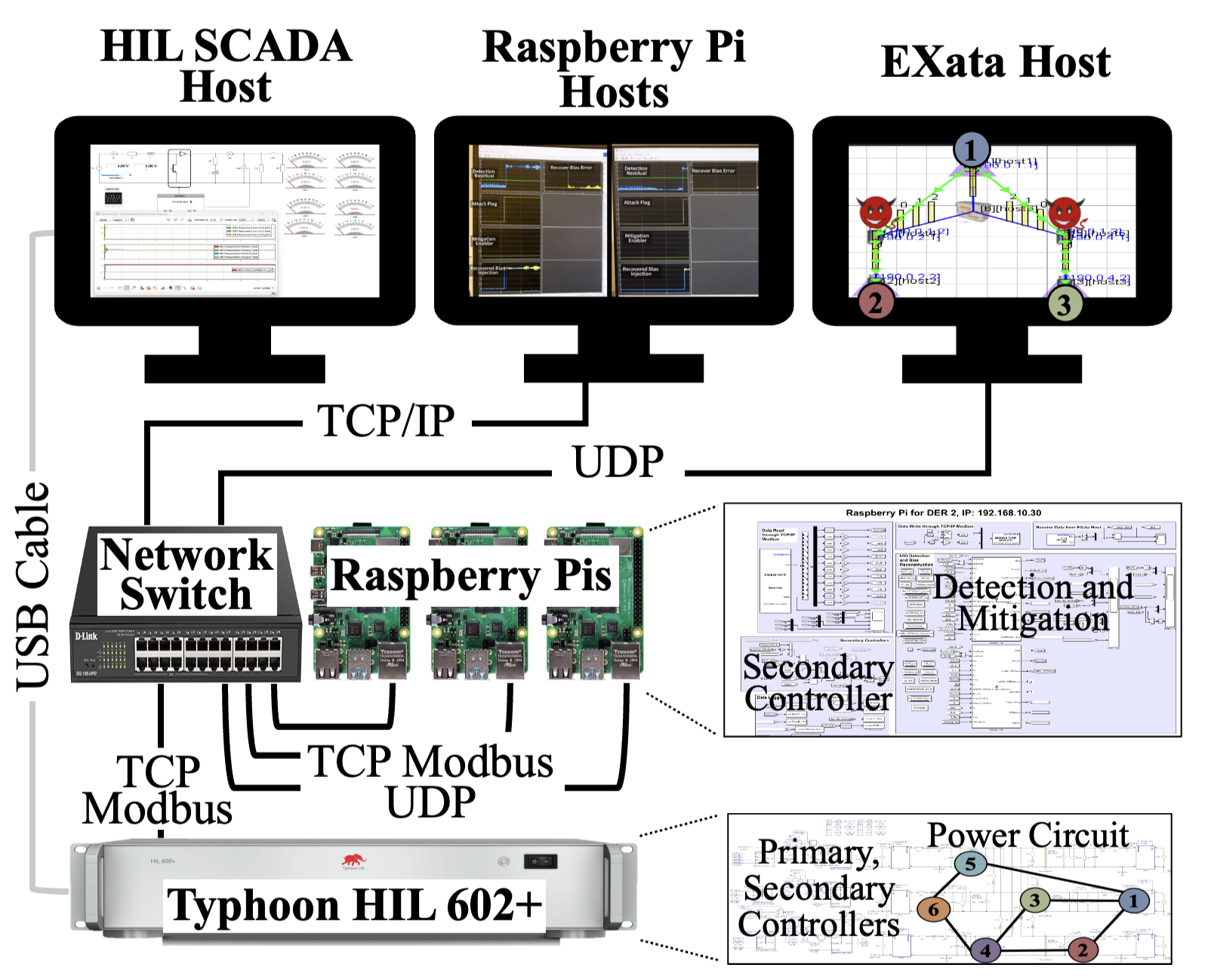}}
  \subfigure[Demonstration of SFDIA implementation in EXata]{\includegraphics[width=8cm]{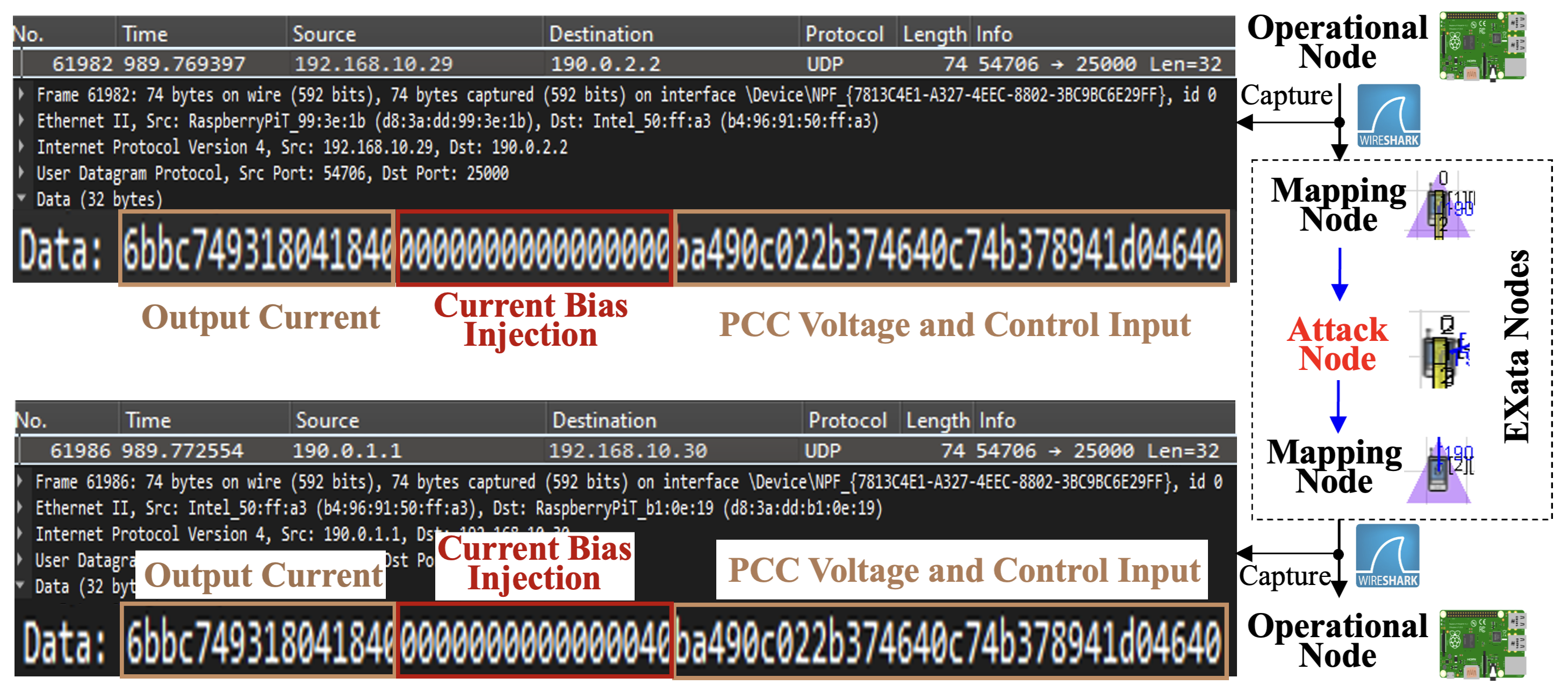}}
  \captionsetup{font={footnotesize}}
  \caption{\color{blue}Sub-figures (a) and (b) show the setup of a 6-DER HIL microgrid in CAPS at University of Sheffield, where the microgrid circuit, primary controllers, and three secondary controllers are implemented in Typhoon HIL 602+, the remaining three secondary controllers and proposed detection-triggered mitigation schemes are run in raspberry Pis, and the data interaction between raspberry Pis are accomplished in EXata network modeling software. Sub-figure (c) demonstrates the SFDIA implementation in EXata. In particular, by interfacing operational nodes, i.e., raspberry pis, with EXata nodes, the real-time data flows between operational nodes will be mapped into EXata. Through utilising the data modification function provided by EXata's cyber libraries, arbitrary bias injections can be introduced into DER's current measurements.}\label{fig:ExperimentSetup}
\end{figure}

{\color{blue}\subsection{Experimental Validations in a 6-DER HIL Testbed}
This subsection validates the proposed method's robustness against load variations and RES voltage fluctuations, the proposed method's superiority compared with the most related slope-based bias reconstruction method \cite{jin2022distributed}, and the proposed method's effectiveness against multiple discontinuous SFDIAs as well as its lightweight computation burden in a HIL microgrid testbed setup at Control and Power Systems (CAPS) lab of University of Sheffield.

Before showcasing the results, we will briefly introduce the HIL microgrid testbed. As shown in Fig. \ref{fig:ExperimentSetup}, the testbed includes three key components, i.e., Typhoon HIL 602+, raspberry pis, and EXata network modelling simulator. The 6-DER microgrid with the topology as shown in the last column of Fig. \ref{fig:CurrentSensorDeploymentDemonstration} is established and simulated in Typhoon HIL 602+, where the electrical parameters are set the same as in \cite{9815319} and additional current sensors are deployed following the cost-effective strategy presented in Algorithm \ref{Alg: current sensor deployment}. Besides simulating the power circuit, HIL 602+ also implements primary controllers and half of secondary controllers. The remaining secondary controllers and proposed detection-triggered mitigation method are embedded into 3 raspberry pis, which interact with HIL 602+ through TCP Modbus every 10ms. Two UDP data flows between raspberry pis are mapped into EXata via external nodel emulation mode, and realistic data modification events can be introduced against mapped data flows utilising the advanced cyber library provided by EXata. Three upper hosts are needed to run in parallel with the testbed to monitor and record the real-time data from HIL 602+ and raspberry pis through USB and Ethernet cables, respectively. In the experiments conducted in the HIL microgrid testbed, the secondary controllers are activated at $t=2$s, followed by the initialisation of UIO-based detectors at $t=4$s. 

\begin{figure}[!h]
  \centering
  \includegraphics[width=9cm]{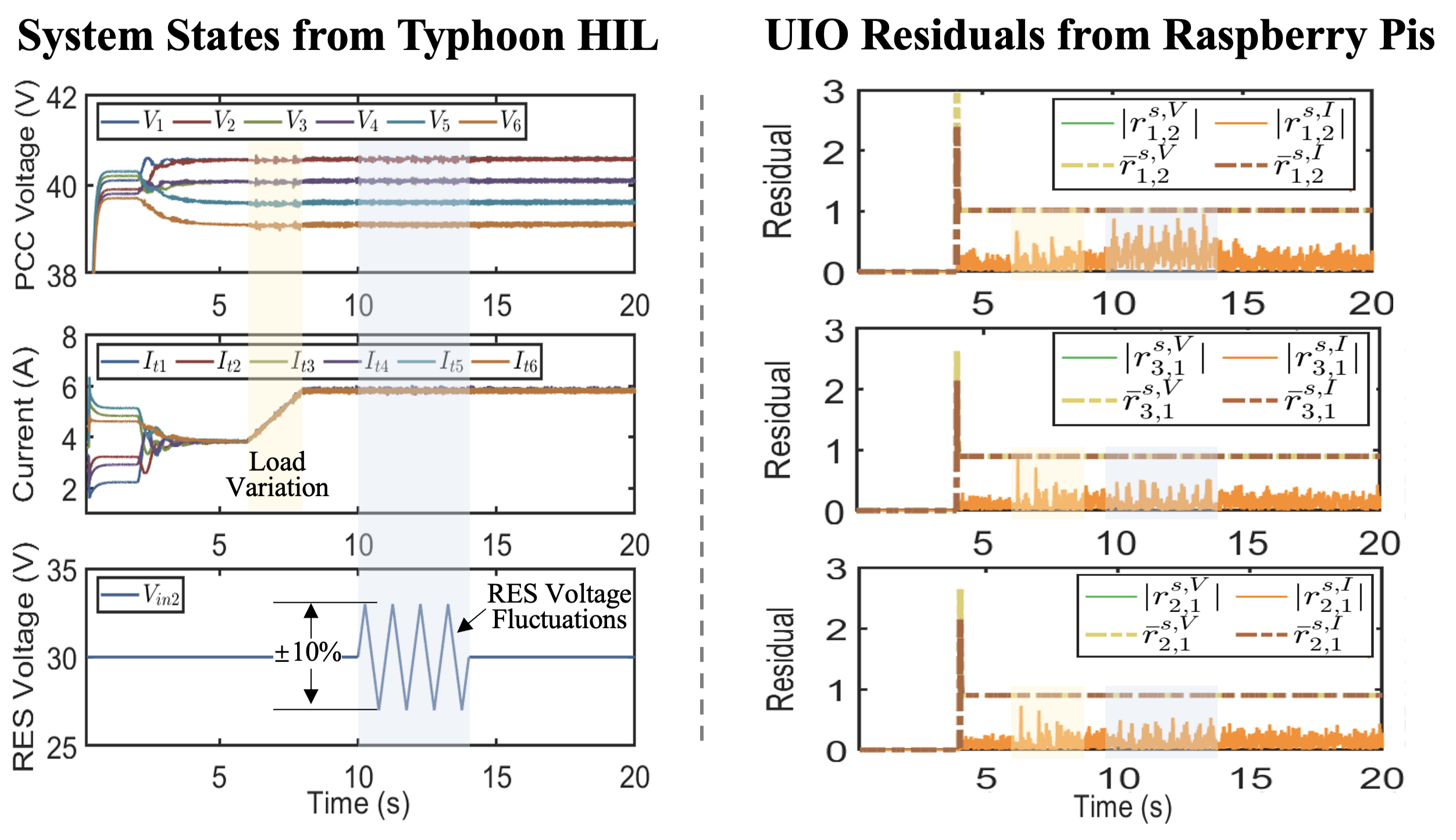}
  \captionsetup{font={footnotesize}}
  \caption{\color{blue}This figure demonstrates the robustness of UIO residuals under load variations and RES voltage fluctuations.}\label{fig:Robustness_against_Daily_Operations}
  \vspace{-10pt}
\end{figure}

\subsubsection{Robustness under Load Variation and RES Voltage Fluctuation} 
Two daily events including current load increase within all DERs and $10\%$ RES voltage fluctuations in DER 2 are introduced within $t \in [6,8]$s and $t \in [10, 14]$s, respectively. 
As shown in Fig. \ref{fig:Robustness_against_Daily_Operations}, the introduced events will result in changes on PCC voltages and output currents, and slightly alter the residuals of three UIOs deployed in raspberry pis, denoted by $\cb{r}_{1,2}^s, \cb{r}_{3,1}^s, \cb{r}_{2,1}^s$, which are still bounded by the appropriately chosen detection thresholds. 

\begin{figure*}[!h]
  \centering
  \includegraphics[width=19cm]{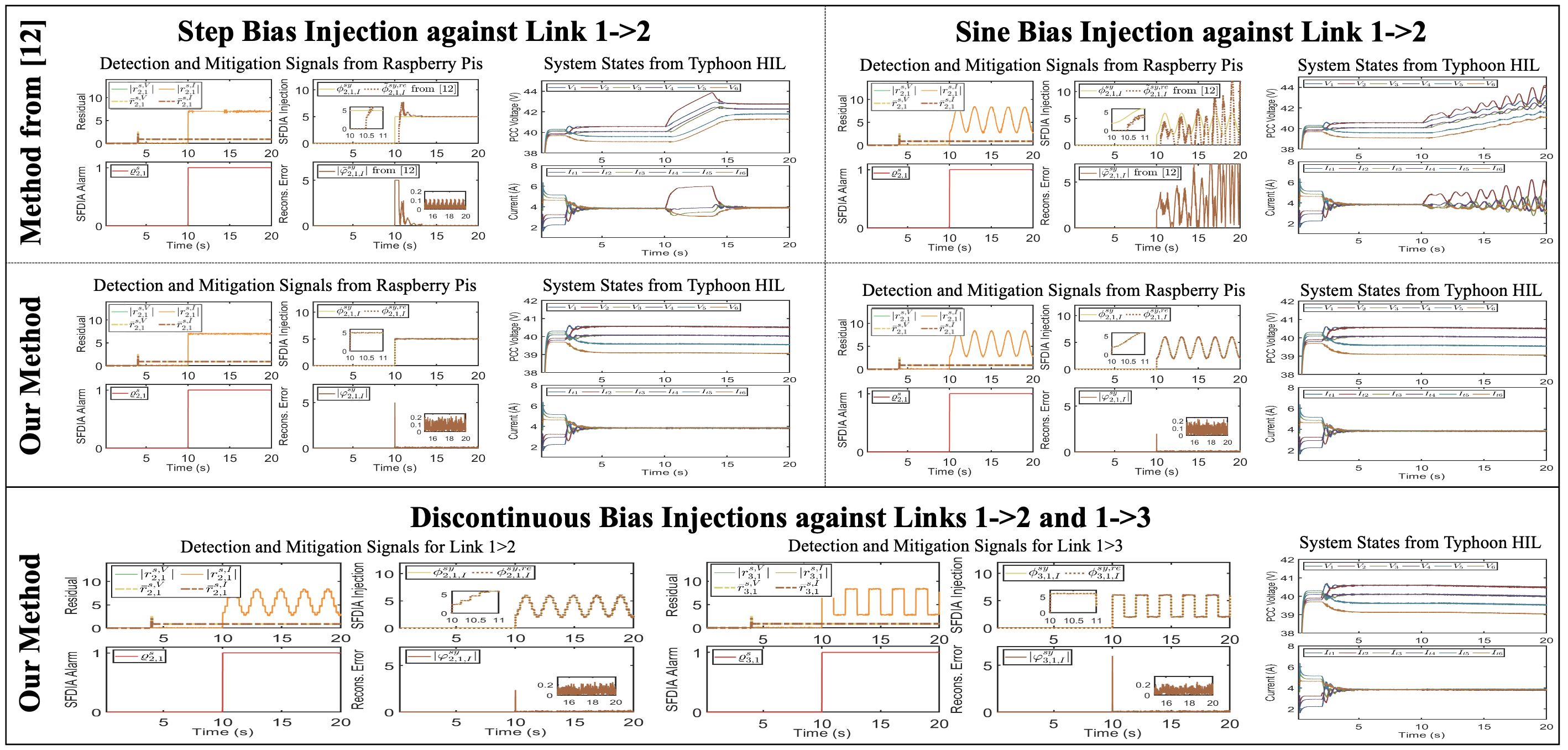}
  \captionsetup{font={footnotesize}}
  \caption{\color{blue}The figure demonstrates the superiority of the proposed method compared with the slope-based bias reconstruction method from \cite{jin2022distributed} (the first two rows) and its effectiveness against multiple discontinuous SFDIAs (the last row) based on the HIL microgrid testbed. }\label{fig:ComparativeStudyandEffectivenessValidation}
  \vspace{-10pt}
\end{figure*}

\subsubsection{Superiority compared with the Slope-based Bias Reconstruction Method \cite{jin2022distributed}} 
The slope-based method \cite{jin2022distributed} was proposed to reconstruct the injected bias by observing the resulted voltage changing slope, which was designed to be effective only for single constant injection. Therefore, this study chooses to compare the reconstruction and mitigation performance of our and slope-based methods when communication link $(2,1)$ is subject to step and sine bias injections at $t=10$s. The time window during which the voltage changing slope is observed is selected as $50$ time steps to eliminate the impacts of voltage harmonics. According to the first two rows' results of Fig. \ref{fig:ComparativeStudyandEffectivenessValidation}, in the presence of step bias injection, the slope-based method can accurately reconstruct the constant bias injection after approximately 5s' observation, with the reconstruction error smaller than $0.2$. After integrating the reconstructed bias, load sharing can be reestablished among DERs but the PCC voltages will have about $2$V deviations from normal values. For the same step bias injection, our method is able to accurately reconstruct the bias injection immediately when attack alarm is flagged, with almost the same level of steady-state reconstruction error. The induced attack impact can be completely eliminated. Hence, our method can accurately reconstruct the constant bias injection much quicker than the slope-based method, and therefore achieve more comprehensive mitigation performance. Under sine bias injection, the slope-based method is not able to accurately reconstruct the injected bias as the resulted voltage changing slope varies too fast, and the adopted imprecise bias reconstruction is likely destabilise the microgrid. On the contrary, our method can accurately reconstruct the varying sine bias injection and thoroughly eliminate the attack impact. Therefore, our method effectively supplements the slope-based method's limitation in addressing the non-constant bias injection and thus has wider application scenarios. For demonstration purpose, we have uploaded a video\footnote{\url{https://www.youtube.com/watch?v=cv2xW-nL2Og}} to showcase the effectiveness of proposed method when links $(2,1)$ and $(3,1)$ are both subject to step bias injections.

\begin{figure}[!h]
  \centering
  \includegraphics[width=9cm]{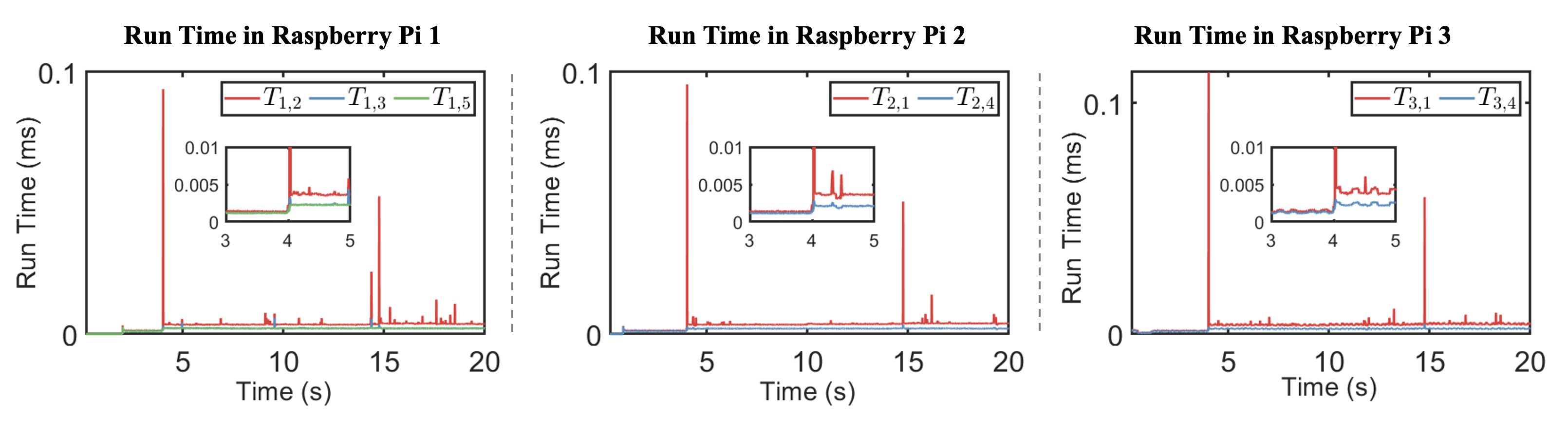}
  \captionsetup{font={footnotesize}}
  \caption{\color{blue}This figure shows the computation time of the proposed detection-triggered impact mitigatin scheme in raspberry pis.}\label{fig:RunTimeinRaspberryPis}
  \vspace{-10pt}
\end{figure}

\subsubsection{Effectiveness against Multiple Discontinuous SFDIAs and Lightweight Computation Burden} Besides the single continuous bias injection, the effectiveness of the proposed method against multiple discontinuous bias injections are also validated as shown in the last row of Fig. \ref{fig:ComparativeStudyandEffectivenessValidation}. When discontinuous sine and rectangle biases are injected into links $(2,1)$ and $(3,1)$, respectively, they can be reconstructed timely and accurately from the residuals generated by UIO$_{2,1}^s$ and UIO$_{3,1}^s$ that are deployed in raspberry pis 2 and 3. The discontinuous bias injection will not affect the reconstruction accuracy and the steady-state reconstruction errors can be always bounded by $0.3$. Moreover, the computation time of the proposed detection-triggered mitigation scheme in 3 raspberry pis are recorded and shown in Fig. \ref{fig:RunTimeinRaspberryPis}. In most cases, the computation time after activating UIOs will not exceed $0.005$ms, and the computation time can increase to $0.1$ms rarely, which, however, is still far smaller than the communication time step $10$ms and thus is negligible. In light of these results, it is reasonable to claim that the proposed detection-triggered impact mitigation scheme is lightweight enough to run in parallel with the secondary controllers.}

\section{Conclusion}
This paper proposed a novel detection-triggered recursive impact mitigation scheme against the SFDIA on communication links in microgrids, which overcomes the limitation of requiring at least one trustworthy neighboring link and only needs to install extra current sensors. With awareness of the physical interconnections among DERs, the voltage bias injection is observed from the power line current readings, based on which a recursive bias reconstruction scheme can be established to accurately obtain the current bias injection. The attack impact is eliminated by subtracting the reconstructed bias from the compromised communication data. {\color{blue}To reduce the cost of deploying current sensors on power lines, a cost-effective deployment strategy is presented to secure a spanning tree subset of communication links that guarantees the secondary control performance. Extensive simulation studies and experimental validations were conducted to verify the correctness of the derived theoretical results and to validate the effectiveness of the proposed method against single/multiple and continuous/discontinuous SFDIAs in the presence of load variations, DERs plugging-in, and RES voltage fluctuations. Future works include investigating the optimal current sensor deployment considering economic dispatch \cite{10366401,10366403} and applying the proposed defense method to load frequency \cite{10124159} and wind turbine \cite{zhao2022dual} control.
}

\section*{Acknowledgement}
The authors would like to thank anonymous reviewers very much for their fruitful and insightful suggestions during the revision of this manuscript.

\appendix
\subsection{DER System Parameters}\label{appendix: system parameters}
{\small\begin{align}\label{eq:system parameters}
{A}_{ii}&=\left[ 
\begin{array}{cc}
{-\frac{1}{Z_{Li}C_{ti}}-\sum\limits_{j\in\mathcal{N}_i^{el}}\frac{1}{C_{ti}R_{ij}}} & \frac{1}{C_{ti}} \\
-\frac{1}{L_{ti}} & -\frac{R_{ti}}{L_{ti}}  \\
\end{array} 
\right], {\cb{b}}_i&=\left[ \begin{array}{c}
0 \\
\frac{1}{L_{ti}}
\end{array}
\right] \nonumber \\
{\cb{m}}_i&=\left[ \begin{array}{c}
-\frac{1}{C_{ti}} \\
0 
\end{array}
\right]
\end{align}}

{\small\begin{align}\label{eq: discreted system parameters}
   &{A}_{ii}^d=e^{{A}_{ii}T_{samp}}, {Y}_{ii}^d=({A}_{ii})^{-1}({A}_{ii}^d-{\rm I}^2), \nonumber \\
   &{\cb{b}}_i^d={Y}_{ii}^d{\cb{b}}_i, {\cb{m}}_i^d={Y}_{ii}^d{\cb{m}}_i.
\end{align}}



\subsection{Proof of Proposition \ref{Pros: Bias Reconstruction under SFDIA}} \label{appendix: proof of proposition 1}

\begin{Proof}
    Rewriting the reconstruction dynamics \eqref{eq: SFDIA bias reconstruction} for $k \ge k_{i,j}^{sa}$ utilising its eigenvalue $\eta_j^s$ as 
    \begin{align}\label{eq: SFDIA reconstruction dynamics}
        &\phi_{i,j,I}^{sy,re} (k+1) = \eta_j^s \phi_{i,j,I}^{sy,re} (k) + (\tilde{\cb{t}}_{j2}^s)^{\rm T}T_j^sA_{jj}^d\cb{\kappa}\phi_{i,j,V}^{sy,ob}(k) + \nonumber \\
        & + (\tilde{\cb{t}}_{j2}^s)^{\rm T}\Big(-\cb{t}_{j1}^s\phi_{i,j,V}^{sy,ob}(k+1)
        + \cb{\psi}_{i,j}^{sa}(k) \Big).
    \end{align}Following the same procedure, the original residual dynamics \eqref{eq: Res Calculation3} for $k \ge k_{i,j}^{sa}$ can be reformulated as
    \begin{align}\label{eq: SFDIA original dynamics}
        &\phi_{i,j,I}^{sy} (k+1) = \eta_j^s \phi_{i,j,I}^{sy} (k) + (\tilde{\cb{t}}_{j2}^s)^{\rm T}T_j^sA_{jj}^d\cb{\kappa}\phi_{i,j,V}^{sy}(k) + \nonumber \\
        & + (\tilde{\cb{t}}_{j2}^s)^{\rm T}\Big(-\cb{t}_{j1}^s\phi_{i,j,V}^{sy}(k+1) +
        \cb{\psi}_{i,j}^{sa}(k) + \cb{\chi}_{i,j}^{s}(k) \Big ).
    \end{align}
    
    Thus, the dynamics of reconstruction error for $k > k_{i,j}^{sa}$ can be obtained by subtracting \eqref{eq: SFDIA reconstruction dynamics} from \eqref{eq: SFDIA original dynamics} as
    \begin{align}
        &\varphi_{i,j,I}^{sy}(k+1) = \eta_j^s \varphi_{i,j,I}^{sy}(k) + (\tilde{\cb{t}}_{j2}^s)^{\rm T}T_j^sA_{jj}^d\cb{\kappa} \hat{\phi}_{i,j,V}^{sy,err}(k) + \nonumber \\
        & + (\tilde{\cb{t}}_{j2}^s)^{\rm T}\Big(-\cb{t}_{j1}^s\hat{\phi}_{i,j,V}^{sy,err}(k+1) 
        + \cb{\chi}_{i,j}^{s}(k) \Big )
    \end{align}where the voltage bias's estimation error $\hat{\phi}_{i,j,V}^{sy,err}(k) = {\phi}_{i,j,V}^{sy}(k) - {\phi}_{i,j,V}^{sy,ob}(k)$ can be approximated as zero for $k\ge k_{i,j}^{sa}$
    Thus, by direct calculation, the steady-state current bias reconstruction error after setting the initial bias injection values as zeros can be obtained as
    {\small\begin{align}\label{eq: steady-state bias reconstruction error expression}
        \varphi_{i,j,I}^{sy}(\infty) = (\eta_j^s &)^{\infty} \Big(\phi_{i,j,I}^{sy}(k_{i,j}^{sa}) 
        \Big) + \nonumber\\ 
        & + \sum_{l=0}^{\infty} (\eta_j^s)^l (\tilde{\cb{t}}_{j2}^s)^{\rm T} \cb{\chi}_{i,j}^{s}(l+k_{i,j}^{sa}).
    \end{align}}
    Since $|\eta_j^s| < 1 $ and $|\cb{\chi}_{i,j}^{s} (k)| \le \bar{\cb{\chi}}_{i,j}^{s}$, the bound of $|\varphi_{i,j,I}^{sy}(\infty)|$ can be calculated as
    {\small\begin{align}
        |\varphi_{i,j,I}^{sy}(\infty)| \le \sum_{l=0}^{\infty} (\eta_j^s)^l (|\tilde{\cb{t}}_{j2}^s|)^{\rm T} \bar{\cb{\chi}}_{i,j}^{s} = \frac{(|\tilde{\cb{t}}_{j2}^s|)^{\rm T} \bar{\cb{\chi}}_{i,j}^{s}}{1-\eta_j^s},
    \end{align}}which completes the proof.
\end{Proof}

\subsection{Proof of Proposition \ref{Pros: Stability Condition Analyse}}\label{Pros: Proof of Pros 2}
\begin{Proof}
According to \eqref{eq: UIOSec_2}, we know that $\cb{t}_{j1}^s = \zeta_j^{md} \cb{t}_{j2}^s$, where $\zeta_j^{md} = -\frac{m_{j2}^d}{m_{j1}^d}$ with $\cb{m}_j^d = [m_{j1}^d, m_{j2}^d]^{\rm T}$. Thus, we have
\begin{align}\label{eq: middle result of stability condition}
    \eta_j^s = (\tilde{\cb{t}}_{j2}^s)^{\rm T}T_j^sA_{jj}^d\cb{\iota} &= [(\tilde{\cb{t}}_{j2}^s)^{\rm T}\cb{t}_{j1}^s, 1]A_{jj}^d\cb{\iota} \nonumber \\
    &= [\zeta_j^{md},1]A_{jj}^d\cb{\iota},
\end{align}which implies that $\eta_j^s$ is irrelevant with the UIO matrix $T_j^s$ but is entirely determined by discrete-time system parameters $A_{jj}^d$ and $\cb{m}_j^d$. According to \eqref{eq: discreted system parameters}, the derivation of $A_{jj}^d$ and $\cb{m}_j^d$ from $A_{jj}$ and $\cb{m}_j$ is closely related to the intractable matrix exponential term $e^{A_{jj}T_{samp}}$. To obtain the insight behind \eqref{eq: middle result of stability condition}, the second-degree Taylor expansion for the approximation of $e^{A_{jj}T_{samp}}$ is used for the subsequent analysis, i.e.,
{\begin{align}\label{eq: middle result of stability condition2}
    e^{A_{jj}T_{samp}} = \sum_{l=0}^{\infty} \frac{1}{l!}(A_{jj}T_{samp})^l \approx {\rm I} + A_{jj}T_{samp} + \frac{1}{2}(A_{jj}T_{samp})^2,
\end{align}}where the higher-degree Taylor expansion term $\sum_{l=3}^{\infty} \frac{1}{l!}(A_{jj}T_{samp})^l$ is sufficiently small since $|a_{jj}^{rc}T_{samp}| < 1, \forall r,c \in \{1,2\}$. Based on \eqref{eq: discreted system parameters} and \eqref{eq: middle result of stability condition2}, \eqref{eq: middle result of stability condition} can be approximated as
{\begin{align}\label{eq: middle result 2 of stability condition}
    \eta_j^s &\approx 1+ a_{jj}^{22}T_{samp} + \frac{1}{2}\big((a_{jj}^{22})^2+a_{jj}^{12}a_{jj}^{21}\big)T_{samp}^2+\nonumber \\
    & - \frac{a_{jj}^{12}a_{jj}^{21}T_{samp}^2\big(2+(a_{jj}^{11}+a_{jj}^{22})T_{samp}\big)}{2(2+a_{jj}^{11}T_{samp})}.
\end{align}}Since $a_{jj}^{11} \approx a_{jj}^{22}$, we have $\frac{\big(2+(a_{jj}^{11}+a_{jj}^{22})T_{samp}\big)}{2+a_{jj}^{11}T_{samp}} \approx 1$, under which \eqref{eq: middle result 2 of stability condition} can be further approximated as
\begin{align}\label{eq: middle result 3 of stability condition}
    \eta_j^s &\approx 1+ a_{jj}^{22}T_{samp} + \frac{1}{2}(a_{jj}^{22}T_{samp})^2.
\end{align}The last term in \eqref{eq: middle result 3 of stability condition} is significantly smaller than the previous term due to $|a_{jj}^{22}T_{samp}| \ll 1$, and thus \eqref{eq: approximation of reconstruction eigenvalue} can be obtained after ignoring the last term. With $a_{jj}^{22}<0$, it can be easily inferred that \eqref{eq: SFDIA bias reconstruction stability condition} always holds, which completes the proof.

\end{Proof}


\bibliographystyle{IEEEtran}
\tiny
\bibliography{root}
\end{spacing}

\end{document}